\documentstyle[12pt,epsf]{article}

\renewcommand{\thefootnote}{\fnsymbol{footnote}}
\newlength{\pubnumber} 

\catcode`\@=11
\@addtoreset{equation}{section}

\def\section{\@startsection{section}{1}{\z@}{3.5ex plus 1ex minus .2ex}
 {2.3ex plus .2ex}{\large\bf}}
\def\subsection{\@startsection{subsection}{2}{\z@}{2.3ex plus .2ex}
 {2.3ex plus .2ex}{\bf}}

\begin{document} 
\begin{titlepage}
\samepage{
\setcounter{page}{1}
\rightline{LTH--779}
\rightline{BU-HEPP 07-07}
\rightline{CASPER 07-03}
\rightline{\tt hep-th/}
\vfill
\begin{center}
 {\Large \bf  Quasi--realistic heterotic--string models\\
with vanishing one--loop cosmological constant \\and perturbatively
broken supersymmetry? \\}
\vfill
 \vfill
 {\large  Gerald B.\  Cleaver\footnote{
   E-mail address: Gerald{\underline{\phantom{B}}}Cleaver@baylor.edu}$^1$,\\
Alon E.\  Faraggi\footnote{
        E-mail address: faraggi@amtp.liv.ac.uk}$^2$,
Elisa Manno\footnote{
	E-mail address: Elisa.Manno@liv.ac.uk}$^2$
and
Cristina Timirgaziu$\setcounter{footnote}{3}$\footnote{
        E-mail address: timirgaz@amtp.liv.ac.uk}$^2$\\
}
\vspace{.12in}
 {\it $^{1}$ CASPER, Department of Physics, Baylor University,
            Waco, TX, 76798-7316}\\
{\it  $^{2}$ Department of Mathematical Sciences,
		University of Liverpool,
                Liverpool L69 7ZL}
\end{center}
\vfill
\begin{abstract}
Quasi--realistic string models in the free fermionic formulation typically
contain an anomalous $U(1)$, which gives rise to a Fayet--Iliopoulos $D$--term
that breaks supersymmetry at the one--loop level in
string perturbation theory. 
Supersymmetry is traditionally restored by imposing $F$-- and $D$--flatness
on the vacuum. By employing the standard analysis of
flat directions 
we present a quasi--realistic three generation string model
in which stringent $F$-- and $D$-- flat solution do not appear to exist
to all orders in the superpotential. 
We speculate that this result is indicative of the non--existence 
of supersymmetric flat F-- and D--solutions in this model. We 
provide some arguments in support of this scenario and discuss
its potential implications. 
Bose--Fermi degeneracy of the string
spectrum implies that the one--loop partition function and hence the one--loop
cosmological constant vanishes in the model. If our assertion is correct,
this model may represent the
first known example with vanishing cosmological constant and perturbatively
broken supersymmetry. 
We discuss the distinctive 
properties of the internal free fermion boundary conditions that
may correspond to a large set of models that share these properties. 
The geometrical moduli in this class of models are fixed due to 
asymmetric boundary conditions, whereas absence of supersymmetric
flat directions would imply that the supersymmetric moduli are fixed
as well and the dilaton may be fixed by hidden sector nonperturbative 
effects.

\end{abstract}
\smallskip}
\end{titlepage}

\renewcommand{\thefootnote}{\arabic{footnote}}
\setcounter{footnote}{0}

\def\l{\label}
\def\beq{\begin{equation}}
\def\eeq{\end{equation}}
\def\beqn{\begin{eqnarray}}
\def\eeqn{\end{eqnarray}}
\def\nolabel{\nonumber }

\def\ie{{\it i.e.}}
\def\eg{{\it e.g.}}
\def\half{{\textstyle{1\over 2}}}
\def\third{{\textstyle {1\over3}}}
\def\quarter{{\textstyle {1\over4}}}
\def\tenth{{\textstyle {1\over{10}}}}
\def\m{{\tt -}}
\def\p{{\tt +}}

\def\slash#1{#1\hskip-6pt/\hskip6pt}
\def\slk{\slash{k}}
\def\GeV{\,{\rm GeV}}
\def\TeV{\,{\rm TeV}}
\def\y{\,{\rm y}}
\def\SM{Standard-Model }
\def\SUSY{supersymmetry }
\def\SSSM{supersymmetric standard model}
\def\vev#1{\left\langle #1\right\rangle}
\def\l{\langle}
\def\r{\rangle}

\def\Htw{{\tilde H}}
\def\chibar{{\overline{\chi}}}
\def\qbar{{\overline{q}}}
\def\ibar{{\overline{\imath}}}
\def\jbar{{\overline{\jmath}}}
\def\Hbar{{\overline{H}}}
\def\Qbar{{\overline{Q}}}
\def\abar{{\overline{a}}}
\def\alphabar{{\overline{\alpha}}}
\def\betabar{{\overline{\beta}}}
\def\tautwo{{ \tau_2 }}
\def\thetatwo{{ \vartheta_2 }}
\def\thetathree{{ \vartheta_3 }}
\def\thetafour{{ \vartheta_4 }}
\def\ttwo{{\vartheta_2}}
\def\tthree{{\vartheta_3}}
\def\tfour{{\vartheta_4}}
\def\ti{{\vartheta_i}}
\def\tj{{\vartheta_j}}
\def\tk{{\vartheta_k}}
\def\calF{{\cal F}}
\def\smallmatrix#1#2#3#4{{ {{#1}~{#2}\choose{#3}~{#4}} }}
\def\ab{{\alpha\beta}}
\def\Minv{{ (M^{-1}_\ab)_{ij} }}
\def\bone{{\bf 1}}
\def\ii{{(i)}}
\def\V{{\bf V}}
\def\b{{\bf b}}
\def\N{{\bf N}}
\def\t#1#2{{ \Theta\left\lbrack \matrix{ {#1}\cr {#2}\cr }\right\rbrack }}
\def\C#1#2{{ C\left\lbrack \matrix{ {#1}\cr {#2}\cr }\right\rbrack }}
\def\tp#1#2{{ \Theta'\left\lbrack \matrix{ {#1}\cr {#2}\cr }\right\rbrack }}
\def\tpp#1#2{{ \Theta''\left\lbrack \matrix{ {#1}\cr {#2}\cr }\right\rbrack }}
\def\l{\langle}
\def\r{\rangle}


\def\inbar{\,\vrule height1.5ex width.4pt depth0pt}

\def\IC{\relax\hbox{$\inbar\kern-.3em{\rm C}$}}
\def\IQ{\relax\hbox{$\inbar\kern-.3em{\rm Q}$}}
\def\IR{\relax{\rm I\kern-.18em R}}
 \font\cmss=cmss10 \font\cmsss=cmss10 at 7pt
\def\IZ{\relax\ifmmode\mathchoice
 {\hbox{\cmss Z\kern-.4em Z}}{\hbox{\cmss Z\kern-.4em Z}}
 {\lower.9pt\hbox{\cmsss Z\kern-.4em Z}}
 {\lower1.2pt\hbox{\cmsss Z\kern-.4em Z}}\else{\cmss Z\kern-.4em Z}\fi}

\def\AEF{A.E. Faraggi}
\def\NPB#1#2#3{{Nucl.\ Phys.}\/ {B \bf #1} (#2) #3}
\def\PLB#1#2#3{{Phys.\ Lett.}\/ {B \bf #1} (#2) #3}
\def\PRD#1#2#3{{Phys.\ Rev.}\/ {D \bf #1} (#2) #3}
\def\PRL#1#2#3{{Phys.\ Rev.\ Lett.}\/ {\bf #1} (#2) #3}
\def\PRP#1#2#3{{Phys.\ Rep.}\/ {\bf#1} (#2) #3}
\def\MODA#1#2#3{{Mod.\ Phys.\ Lett.}\/ {\bf A#1} (#2) #3}
\def\IJMP#1#2#3{{Int.\ J.\ Mod.\ Phys.}\/ {A \bf #1} (#2) #3}
\def\nuvc#1#2#3{{Nuovo Cimento}\/ {\bf #1A} (#2) #3}
\def\JHEP#1#2#3{{JHEP} {\textbf #1}, (#2) #3}
\def\EPJC#1#2#3{{Eur.\ Phys.\ Jour.}\/ {C \bf #1} (#2) #3}
\def\etal{{\it et al\/}}

\newcommand{\be}{\begin{equation}}
\newcommand{\ee}{\end{equation}}
\newcommand{\ba}{\begin{eqnarray}}
\newcommand{\ea}{\end{eqnarray}}
\hyphenation{su-per-sym-met-ric non-su-per-sym-met-ric}
\hyphenation{space-time-super-sym-met-ric}
\hyphenation{mod-u-lar mod-u-lar--in-var-i-ant}
\def\bda#1{{${\cal{D}}_{#1}$}}
\topmargin=20mm
\setcounter{footnote}{0}
\section{Introduction}
\bigskip

The quasi--realistic heterotic--string models in the free fermionic
formulation, which are related to $Z_2\times Z_2$ orbifold compactifications,
are among the most realistic string models constructed to date.
These models produce a rich variety of three generation models with the 
canonical $SO(10)$ embedding of the Standard Model spectrum, and 
include: the flipped $SU(5)$ string
models \cite{fsu5} (FSU5); the standard--like string models
\cite{fny,eu,top,cfn,cfnw,fmt};
the Pati--Salam string models \cite{alr} (PS); the Left--Right
symmetric string models \cite{cfs} (LRS). 
Many of the issues pertaining to the phenomenology of the Standard Model
and Grand Unification were investigated in the context of these models
\cite{reviewffm}.
Additionally, the free fermionic models produced the first known
string models in which the matter content in the observable Standard Model
charged sector of the effective low energy quantum field theory
consists solely of the Minimal Supersymmetric Standard Model
\cite{cfn,cfnw}.

A common feature of many of the quasi--realistic free fermionic 
heterotic--string models is the existence of an ``anomalous'' $U(1)$
symmetry \cite{Cleaver:1997rk}. The anomalous $U(1)_A$ is broken by the
Green--Schwarz--Dine--Seiberg--Witten mechanism \cite{dsw}
in which a potentially large Fayet--Iliopoulos $D$--term
$\xi$ is generated by the VEV of the dilaton field.
Such a $D$--term would, in general, break supersymmetry, unless
there is a direction $\hat\phi=\sum\alpha_i\phi_i$ in the scalar
potential for which $\sum Q_A^i\vert\alpha_i\vert^2$ is of opposite sign to
$\xi$ and that
is $D$--flat with respect to all the non--anomalous gauge symmetries,
as well as $F$--flat. 
If such a direction
exists, it will acquire a VEV, cancelling the Fayet--Iliopoulos
$\xi$--term, restoring supersymmetry and stabilising the vacuum.
The set of $D$- and $F$-flat constraints is given by
\beqn
&& \langle D_A\rangle=\langle D_\alpha\rangle= 0~;\quad
\langle F_i\equiv
{{\partial W}\over{\partial\eta_i}}\rangle=0~~;\label{dterms}\\
\nonumber\\
&& D_A=\left[K_A+
\sum Q_A^k\vert\chi_k\vert^2+\xi\right]~~;\label{da}\\
&& D_\alpha=\left[K_\alpha+
\sum Q_\alpha^k\vert\chi_k\vert^2\right]~,~\alpha\ne A~~;\label{dalpha}\\
&& \xi={{g^2({\rm Tr} Q_A)}\over{192\pi^2}}M_{\rm Pl}^2~~;
\label{dxi}
\eeqn
where $\chi_k$ are the fields which acquire VEVs of order
$\sqrt\xi$, while the $K$--terms contain fields $\eta_i$
like squarks, sleptons and Higgs bosons whose
VEVs vanish at this scale. $Q_A^k$ and $Q_\alpha^k$ denote the anomalous
and non--anomalous charges,
and $M_{\rm Pl}\approx2\times 10^{18}$ GeV denotes the
reduced Planck mass. The solution ({\it i.e.}\  the choice of fields
with non--vanishing VEVs) to the set of
equations (\ref{dterms})--(\ref{dalpha}),
though nontrivial, is not unique. Therefore in a typical model there exist
a moduli space of solutions to the $F$ and $D$ flatness constraints,
which are supersymmetric and degenerate in energy \cite{moduli}. Much of
the study of the superstring models phenomenology 
(as well as non--string supersymmetric models \cite{savoy})
involves the analysis and classification of these flat directions.
The methods for this analysis in string models have been systematised
in \cite{systematic1,systematic2,cfn,mshsm}.

In general it has been assumed in the past that in a given string model
there should exist a supersymmetric solution to the $F$ and $D$
flatness constraints. The simpler type of solutions utilise only
fields that are singlets of all the non--Abelian groups in a given
model (type I solutions). More involved solutions (type II solutions),
that utilise also non--Abelian fields, have also been considered
\cite{mshsm}, as well as inclusion of non--Abelian fields
in systematic methods of analysis \cite{mshsm}.
The general expectation that a given model admits a supersymmetric 
solution arises from analysis of supersymmetric point quantum field theories.
In these cases it is known that if supersymmetry is preserved at the 
classical level, {\it i.e.} tree--level in perturbation theory, 
then there exist index theorems that forbid supersymmetry breaking 
at the perturbative quantum level \cite{witten1982}.
Therefore in point quantum field theories
supersymmetry breaking
may only be induced by non--perturbative effects \cite{is}.

Recently we constructed string models \cite{fmt} in which the
issues of supersymmetric flat directions merits further investigation.
The aim of ref.\  \cite{fmt} was to build models with a
reduced untwisted Higgs spectrum. This was achieved in ref.\  \cite{fmt}
by imposing asymmetric boundary conditions in a boundary condition
basis vector that does not break the $SO(10)$ symmetry. 
An unforeseen consequence of the Higgs reduction mechanism of ref.\  \cite{fmt}
was the simultaneous projection of untwisted $SO(10)$ singlet fields. 
Subsequently the moduli space of supersymmetric flat
solutions is vastly reduced. In fact, in ref.\  \cite{fmt} it was
concluded that the model under investigation there does not 
contain supersymmetric flat directions that do not break some
of the Standard Model symmetries. Indeed, for that reason,
a phenomenologically viable model with a reduced untwisted Higgs
spectrum was not presented in ref.\  \cite{fmt}.

The question therefore remains whether a phenomenologically
viable model with a reduced untwisted Higgs spectrum
exists. In this paper we explore this question further.
The untwisted Higgs reduction mechanism that we use
here differs from the one of ref.\  \cite{fmt}. 
Here we present a model that utilises boundary 
conditions that are both symmetric and asymmetric
in the basis vectors that break $SO(10)$ to $SO(6)\times SO(4)$,
with respect to two 
of the twisted sectors of the $Z_2\times Z_2$ orbifold.
The consequence is that two of the untwisted Higgs multiplets,
associated with two of the twisted sectors, are projected entirely
from the massless spectrum. As a result, and similar
to the model of ref.\  \cite{fmt}, the string model
contains a single pair of untwisted electroweak Higgs doublets.

In the process of seeking such a model with a phenomenologically
viable supersymmetric flat direction, we arrive in this paper to the unexpected
conclusion that the model may not contain
supersymmetric flat directions at all.
In the least, this model appears to have no $D$-flat directions that can
be proven
to be $F$-flat to all order, other than through order-by-order analysis. 
That is, there does not appear to be any $D$-flat directions with {\it 
stringent} $F$-flatness (as defined in \cite{cfn,cfnw,Cleaver:2007ek}).
In the analysis of the flat directions we include all the fields in the
string model, {\it i.e.} Standard Model singlet states as well as
Standard Model charged states. The model therefore does not contain a
$D$--flat directions that is also stringently $F$--flat to all order
of non--renormalizable terms. 

The model may of course still admit non-stringent flat directions that rely 
on cancellations between superpotential terms. However, past experience
suggests that non--stringent flat directions
can only hold order by order, and are not maintained to all orders
\cite{pastexperience,cfs}.
This is the key difference between the string theory case, in which 
heavy string modes
generate an infinite tower of terms, versus the field theory case in which
heavy modes are not integrated out.
We therefore speculate that in this case supersymmetry is not exact, 
but is in general broken at some order.

If this finding remains true after the entire parameter 
space of possible all-order non--stringent flat
directions has been examined (an undertaking of several years),
we must ask what are the the implications. 
If a model without all-order
$F$-flatness were to be found, then supersymmetry would remain broken by the
Fayet--Iliopoulos term at a finite order, which is generated at the one--loop 
level in string perturbation theory, rather than be cancelled by a $D$-flat 
direction with anomalous charge. If so, then this would imply, although 
supersymmetry is unbroken at the classical level and the string spectrum is 
Bose--Fermi degenerate, that supersymmetry may be broken at the perturbative
quantum level. Nevertheless, since the spectrum is Bose--Fermi degenerate,
the one--loop cosmological constant still vanishes. 

The string model that we present contains three chiral generations,
charged under the Standard Model gauge group and with the canonical
$SO(10)$ embedding of the weak--hypercharge; one pair of untwisted
electroweak Higgs doublets; a cubic level top--quark Yukawa coupling.
The string model therefore shares some of the phenomenological
characteristics of the quasi--realistic free fermionic string models.
It may therefore represent an example of a quasi--realistic string model,
with vanishing one--loop cosmological constant and perturbatively
broken supersymmetry. 

Our paper is organised as follows: 
in section \ref{review} we
review some aspects of the free fermionic formalism and 
in section \ref{stringent} we elaborate
on stringent flat directions. Then in
section \ref{stringmodelsection} we present the string model.
In section \ref{flatdir} we discuss the methodology of flat direction
analysis and the evidence for concluding that supersymmetric (stringent)
flat directions may not exist to all order in the string model of section
\ref{stringmodelsection}. Section \ref{conclude} concludes the paper.

\section{Free Fermionic Models}\label{review}

In this section we briefly review the construction and structure of the
free fermionic standard like models.
The notation and further details of the construction of these
models are given elsewhere \cite{fny,eu,nahe,cfn,cfs,fmt}.
In the free fermionic formulation of the heterotic string
in four dimensions \cite{fff} all the world--sheet
degrees of freedom,  required to cancel
the conformal anomaly, are represented in terms of free fermions
propagating on the string world--sheet.
In the light--cone gauge the world--sheet field content consists
of two transverse left-- and right--moving space--time coordinate bosons,
$X_{1,2}^\mu$ and ${\bar X}_{1,2}^\mu$,
and their left--moving fermionic superpartners, $\psi^\mu_{1,2}$,
and additional 62 purely internal
Majorana--Weyl fermions, of which 18 are left--moving
and 44 are right--moving.
The models are constructed by specifying the phases picked by
the world--sheet fermions when transported along the torus
non--contractible loops
\be  f \rightarrow -e^{i\pi\alpha(f)}f, \quad \alpha(f) \in (-1,1] .\ee
Each model corresponds to a particular choice
of fermion phases consistent with modular invariance and is generated
by a set of basis vectors describing the transformation properties
of the 64 world--sheet fermions.
The physical spectrum is obtained by applying the generalised GSO projections.
The low energy effective field theory is obtained by S--matrix elements
between external states \cite{kln}.

The boundary condition basis defining a typical
realistic free fermionic heterotic string model is
constructed in two stages.
The first stage consists of the NAHE set,
which is a set of five boundary condition basis vectors,
$\{{\bf1},S,b_1,b_2,b_3\}$ \cite{costas,nahe}.
The gauge group, after imposing the GSO projections induced
by the NAHE set, is $SO(10)\times SO(6)^3\times E_8$,
with $N=1$ supersymmetry.
The NAHE set divides the internal world--sheet
fermions in the following way: ${\bar\phi}^{1,\cdots,8}$ generate the
hidden $E_8$ gauge group, ${\bar\psi}^{1,\cdots,5}$ generate the $SO(10)$
gauge group, while $\{{\bar y}^{3,\cdots,6},{\bar\eta}^1\}$,
$\{{\bar y}^1,{\bar y}^2,{\bar\omega}^5,{\bar\omega}^6,{\bar\eta}^2\}$ and 
$\{{\bar\omega}^{1,\cdots,4},{\bar\eta}^3\}$ generate the three horizontal
$SO(6)$ symmetries. The left--moving $\{y,\omega\}$ states are divided
to $\{{y}^{3,\cdots,6}\}$,
$\{{y}^1,{y}^2,{\omega}^5,{\omega}^6\}$,
$\{{\omega}^{1,\cdots,4}\}$, while $\chi^{12}$, $\chi^{34}$, $\chi^{56}$
generate the left--moving $N=2$ world--sheet supersymmetry.

The second stage of the
basis construction consists of adding to the
NAHE set three additional boundary condition basis vectors.
These additional basis vectors reduce the number of generations
to three chiral generations, one from each of the sectors $b_1$,
$b_2$ and $b_3$, and simultaneously break the four dimensional
gauge group. The assignment of boundary conditions to
$\{{\bar\psi}^{1,\cdots,5}\}$ breaks $SO(10)$ to one of its subgroups.
Similarly, the hidden $E_8$ symmetry is broken to one of its
subgroups. The flavour $SO(6)^3$ symmetries in the NAHE--based models
are always broken to flavour $U(1)$ symmetries, as the breaking
of these symmetries is correlated with the number of chiral
generations. Three such $U(1)_j$ symmetries are always obtained
in the NAHE based free fermionic models from the subgroup
of the observable $E_8$, which is orthogonal to $SO(10)$.
These are produced by the world--sheet currents ${\bar\eta}^j{\bar\eta}^{j^*}$
($j=1,2,3$), which are part of the Cartan sub--algebra of the
observable $E_8$. Additional unbroken $U(1)$ symmetries, denoted
typically by $U(1)_j$ ($j=4,5,...$), arise by pairing two real
fermions from the sets
$\{{\bar y}^{3,\cdots,6}\}$,
$\{{\bar y}^{1,2},{\bar\omega}^{5,6}\}$ and
$\{{\bar\omega}^{1,\cdots,4}\}$.
The final observable gauge
group depends on the number of such pairings.
Alternatively, a left--moving real fermion from the sets
$\{{ y}^{3,\cdots,6}\}$, $\{{ y}^{1,2},{\omega}^{5,6}\}$ and
$\{{\omega}^{1,\cdots,4}\}$ may be paired with its respective
right--moving real fermion to form an Ising model operator,
in which case the rank of the right--moving gauge group is reduced by one.
The reduction of untwisted
electroweak Higgs doublets crucially depends on the pairings
of the left-- and right--moving fermions from the set
$\{y,\omega|{\bar y},{\bar\omega}\}^{1\cdots6}$.

Subsequent to constructing the basis vectors and extracting the massless
spectrum, the analysis of the free fermionic models proceeds by
calculating the superpotential. The cubic and higher-order terms in
the superpotential are obtained by evaluating the correlators
\beq
A_N\sim \langle V_1^fV_2^fV_3^b\cdots V_N^b\rangle,
\label{supterms}
\eeq
where $V_i^f$ $(V_i^b)$ are the fermionic (scalar) components
of the vertex operators, using the rules given in~\cite{kln}.
Typically, one of the $U(1)$ factors in the free-fermion models is anomalous
and generates a Fayet--Iliopoulos term which breaks supersymmetry
at the Planck scale \cite{dsw}. A supersymmetric vacuum is obtained by
assigning non--trivial VEVs to a set of Standard Model singlet
fields in the massless string spectrum along $F$ and $D$--flat directions.
Some of these fields will appear in the nonrenormalizable terms
(\ref{supterms}), leading to
effective operators of lower dimension. Their coefficients contain
factors of order ${\cal V} / M{\sim 1/10}$.

\section{Stringent flat directions}\label{stringent}

In general, systematic analysis of simultaneously $D$- and $F$-flat directions 
in anomalous models is a complicated, non-linear 
process\footnote{In ref.\ \cite{Buchmuller:2006ik} it is argued that, in 
addition to flat directions,
isolated special points generically exist in the VEV 
parameter space that are not located along flat directions, but for which all 
$D$- and $F$-terms are nonetheless zero.
Ref. \cite{Buchmuller:2006ik} calls upon the 
proof by Wess and Bagger \cite{Wess:1992cp}
that non-anomalous $D$-terms do not 
actually increase the number of constraints for supersymmetric flat directions 
beyond the $F$-term constraints. However, FI term cancellation, which requires
the existence of one monomial that is $D$-flat for all non-anomalous
symmetries, but that carries the opposite sign to the FI-term in the
anomalous $U(1)_A$, imposes an additional constraint. Thus, without an
anomalous $U(1)$, the system of $D$- and $F$-equations is not over
constraining. In the latter case, once a solution to $F_m = 0$, for all
fields $s^o _m$, is found (to a given order), 
complexified gauge transformations of the fields $s^o_m$, that continue to 
provide a $F_m=0$ solution, can be performed that simultaneously arrange for 
non-anomalous $D$-flatness. Thus, since the $F_m = 0$ equations impose $m$ 
(non-linear) constraints on $m$ fields, there should be at least one
non--trivial non-anomalous $D$-flat solution for any set of fields $s^o_m$.
A parallel proof for non-Abelian field VEVs also exists. (Complications to
these proofs do arise when different scalar fields possess the same gauge
charges.) This reduction in apparent total constraints is possible because
the $F$-term equations constrain gauge invariant polynomials, which also
correspond to non-anomalous $D$-flat directions \cite{Buchmuller:2006ik}.}.
In weakly coupled heterotic string (WCHS) model-building, $F$-flatness of a
specific VEV direction in the low energy effective field theory may be proven
to a given order by cancellation of $F$-term components, only to be lost a
mere one order higher at which cancellation is not found. An exception is
directions with stringent $F$-flatness \cite{cfn,cfnw,Cleaver:2007ek}.
Rather than allowing cancellation between two or more components in an
$F$-term, stringent $F$-flatness requires that each possible component in an
$F$-term have zero vacuum expectation value. 

When only non-Abelian singlet fields acquire VEVs, stringent flatness
implies that two or more singlet fields in a given $F$-term cannot take
on VEVs. For example, in section 4.1, which presents the third through fifth
order superpotential for the model under consideration, the components of
the $F$-term for $\Phi_{45}$ are (through third order):
\beqn
F_{\Phi_{45}} &=& \bar{\Phi}_{46}\bar{\Phi}'_{56}+
\bar{\Phi}'_{46}\bar{\Phi}_{56}.
\label{fflat1}
\eeqn
For stringent $F$-flatness we require not just that $<F_{\Phi_{45}}> = 0$, but 
that each component within is zero, i.e., 
\beqn
<\bar{\Phi}_{46}\bar{\Phi}'_{56}> = 0,\, <\bar{\Phi}'_{46}\bar{\Phi}_{56}> = 0.
\label{fflat2}
\eeqn
Thus, by not allowing cancellation between components in a given $F$-term, 
stringent $F$-flatness imposes stronger constraints than generic $F$-flatness, 
but requires significantly less fine-tuning between the VEVs of fields.

The net effect of all stringent $F$-constraints on a given superpotential term 
is that at least two fields in the term must not take on VEVs. This condition 
can be relaxed when non-Abelian fields acquire VEVs. Self-cancellation of a 
single component in a given $F$-term is possible between various VEVs within a 
given non-Abelian representation. Self-cancellation was discussed
in \cite{cfnw} for $SU(2)$ and $SO(2n)$ states.

A given set of stringent $F$ flatness constraints are not independent and 
solutions to a set can be expressed in the language of Boolean algebra (logic) 
and applied as constraints to linear combinations of $D$-flat basis
directions.The Boolean algebra language makes clear that the effect
of stringent $F$-flat constraints is strongest for low order superpotential
terms and lessens with increasing order. In particular, for the model
presented herein, stringent flatness is extremely constraining on VEVs of
the reduced number of (untwisted) singlet fields appearing in the third
through fifth order superpotential, in comparison to its constraints on the
larger number of singlets in the model of \cite{fmt}.  

One might imagine that 
stringent $F$-flatness constraints requires order-by-order testing of 
superpotential terms.
This is, in fact, not necessary. All-order stringent $F$-flatness can actually
be proven or disproven by examining only a small finite set of possible dangerous
(i.e., $F$-flatness breaking) superpotential terms. 
Through a process such as matrix 
singular value decomposition (SVD)\footnote{A SVD fortran subroutine is provided in
\cite{numrec}.},
a finite set of superpotential terms can be constructed that generates all possible 
dangerous superpotential terms for a specific $D$-flat direction. This basis of
gauge-invariants can always be formed with particular attributes: 
(1) each basis element term contains at most one unVEVed field (since to threaten
$F$-flatness, a gauge-invariant term, necessarily without anomalous charge, 
can contain no more than one unVEVed field);
(2) there is at most one basis term for each unVEVed field in the model; and
(3) when an unVEVed field appears in a basis term, it appears only to the first power.
The SVD process generated a possibly threading basis of superpotential terms for several
models (see for example \cite{cfn,cfnw,cfs,mshsm,Cleaver:2000sc,Cleaver:2002ps,
Perkins:2005zh}).

To appear in a string-based superpotential, a gauge invariant term must also  
follow Ramond-Neveu-Schwarz worldsheet charge conservation rules.
For free fermionic models these rules were generalized from finite order in \cite{kln,rt}
to all-order in \cite{mshsm}. The generic all order rules can be applied to systematically
determine if any product of SVD-generated $F$-flatness threatening superpotential basis
elements survive in the corresponding string-generated superpotential. If none survive,
then $F$-flatness is proven to all finite order. This technique has been used to prove
$F$-flatness to all finite order for various directions in several models 
\cite{cfn,cfnw,cfs,mshsm,Cleaver:2000sc,Cleaver:2002ps,Perkins:2005zh}.
Alternately, if any terms do survive, the lowest order is determined at which stringent
$F$-flatness is broken.

How should stringent (especially all-order) flat directions be interpreted in 
comparison to general (perhaps finite order) flat directions? All-order 
stringent flat directions contain a minimum
number of VEVs and appear in models 
as the roots of more fine-tuned (generally finite-order) flat directions that
require specific cancellations between $F$-term components. The latter may 
involve cancellations between sets of components of different orders in the 
superpotential. 

All-order stringent flat directions have indeed been
discovered to be such roots 
in all prior free fermionic heterotic models for which we have performed 
systematic flat direction classifications. However, the model presented herein 
appears to lack any stringent flat directions, at least within the expected 
range of VEV parameter space. We have reached this conclusion after employing 
our standard systematic methodology for $D$- and $F$-flat direction analysis.

\section{The String Model}\label{stringmodelsection}

In this section we give the details of our model.
 The boundary condition
basis vectors beyond the NAHE--set and the one--loop GSO projection 
coefficients are shown in eq (\ref{stringmodel}) and eq.
(\ref{phasesmodel1}), respectively.

\beqn
 &\begin{tabular}{c|c|ccc|c|ccc|c}
 ~ & $\psi^\mu$ & $\chi^{12}$ & $\chi^{34}$ & $\chi^{56}$ &
        $\bar{\psi}^{1,...,5} $ &
        $\bar{\eta}^1 $&
        $\bar{\eta}^2 $&
        $\bar{\eta}^3 $&
        $\bar{\phi}^{1,...,8} $ \\
\hline
\hline
  ${\alpha}$  &  0 & 0&0&0 & 1~1~1~0~0 & 1 & 0 & 0 & 1~1~0~0~0~0~0~0 \\
  ${\beta}$   &  0 & 0&0&0 & 1~1~1~0~0 & 0 & 1 & 0 & 0~0~1~1~0~0~0~0 \\
  ${\gamma}$  &  0 & 0&0&0 &
		${1\over2}$~${1\over2}$~${1\over2}$~${1\over2}$~${1\over2}$
	      & ${1\over2}$ & ${1\over2}$ & ${1\over2}$ &
                0~0~0~0~$1\over2$~$1\over2$~${1\over2}$~${1\over2}$ \\
\end{tabular}
   \nonumber\\
   ~  &  ~ \nonumber\\
   ~  &  ~ \nonumber\\
     &\begin{tabular}{c|c|c|c}
 ~&   $y^3{y}^6$
      $y^4{\bar y}^4$
      $y^5{\bar y}^5$
      ${\bar y}^3{\bar y}^6$
  &   $y^1{\omega}^5$
      $y^2{\bar y}^2$
      $\omega^6{\bar\omega}^6$
      ${\bar y}^1{\bar\omega}^5$
  &   $\omega^2{\omega}^4$
      $\omega^1{\bar\omega}^1$
      $\omega^3{\bar\omega}^3$
      ${\bar\omega}^2{\bar\omega}^4$ \\
\hline
\hline
$\alpha$ & 1 ~~~ 0 ~~~ 0 ~~~ 1  & 0 ~~~ 0 ~~~ 1 ~~~ 1  & 0 ~~~ 0 ~~~ 1 ~~~ 1 \\
$\beta$  & 0 ~~~ 0 ~~~ 1 ~~~ 1  & 1 ~~~ 0 ~~~ 0 ~~~ 1  & 0 ~~~ 1 ~~~ 0 ~~~ 1 \\
$\gamma$ & 0 ~~~ 1 ~~~ 0 ~~~ 0  & 0 ~~~ 1 ~~~ 0 ~~~ 0  & 1 ~~~ 0 ~~~ 0 ~~~ 0 \\
\end{tabular}
\label{stringmodel}
\eeqn
With the choice of generalised GSO coefficients:

\begin{equation}
{\bordermatrix{
        &{\bf 1}&  S & &{b_1}&{b_2}&{b_3}& &{\alpha}&{\beta}&{\gamma}\cr
 {\bf 1}&   ~~1 &~~1 & & -1  &  -1 & -1  & &  -1    &  -1   & ~~i   \cr
       S&   ~~1 &~~1 & &~~1  & ~~1 &~~1  & &  -1    &  -1   &  -1   \cr
        &       &    & &     &     &     & &        &       &       \cr
   {b_1}&    -1 & -1 & & -1  &  -1 & -1  & &  -1    &  -1   & ~~i   \cr
   {b_2}&    -1 & -1 & & -1  &  -1 & -1  & &  -1    & ~~1   & ~~i   \cr
   {b_3}&    -1 & -1 & & -1  &  -1 & -1  & & ~~1    &  -1   & ~~1   \cr
	&       &    & &     &     &     & &        &       &       \cr
{\alpha}&    -1 & -1 & & -1  &  -1 &~~1  & & ~~1    & ~~1   & ~~1   \cr
 {\beta}&    -1 & -1 & & -1  & ~~1 & -1  & &  -1    & ~~1   & ~~1   \cr
{\gamma}&    -1 & -1 & &~~1  & ~~1 & -1  & &  -1    &  -1   &  -i   \cr}}
\label{phasesmodel1}
\end{equation}

In matrix (\ref{phasesmodel1}) only the entries above the diagonal are
independent, while those below and on the diagonal are fixed by
the modular invariance constraints. 
Blank lines are inserted to emphasise the division of the free
phases between the different sectors of the realistic
free fermionic models. Thus, the first two lines involve
only the GSO phases of $c{{\{{\bf 1},S\}}\choose a_i}$. The set
$\{{\bf 1},S\}$ generates the $N=4$ model with $S$ being the
space--time supersymmetry generator and therefore the phases
$c{S\choose{a_i}}$ are those that control the space--time supersymmetry
in the superstring models. Similarly, in the free fermionic
models, sectors with periodic and anti--periodic boundary conditions,
of the form of $b_i$, produce the chiral generations.
The phases $c{b_i\choose b_j}$ determine the chirality
of the states from these sectors.

Both the basis vectors $\alpha$ and $\beta$ break the 
$SO(10)$ symmetry to $SO(6)\times SO(4)$ and the basis vector
$\gamma$ breaks it further to $SU(3)\times U(1)_C\times SU(2)\times U(1)_L$.  
The basis vector $\alpha$ is symmetric with respect to the sector
$b_1$ and asymmetric with respect to the sectors $b_2$ and $b_3$, 
whereas the basis vector $\beta$ is symmetric with respect to $b_2$ 
and asymmetric with respect to $b_1$ and $b_3$. As a consequence of these
assignments and of the
string doublet--triplet splitting mechanism \cite{ps}, both the untwisted
Higgs colour triplets and electroweak doublets, with leading coupling
to the matter states from the sectors $b_1$ and $b_2$, are projected
out by the generalised GSO projections. 
At the same time the untwisted colour Higgs triplets 
that couple at leading order to the states from the sector $b_3$ are projected out, 
whereas the untwisted electroweak Higgs doublets
remain in the massless spectrum. Due to the asymmetric boundary 
conditions in the sector $\gamma$ with respect to the sector
$b_3$, the leading Yukawa coupling is that of the up--type quark from the
sector $b_3$ to the untwisted electroweak Higgs doublet \cite{top}.
Hence, the leading Yukawa term is that of the top quark and only its
mass is characterised by the electroweak VEV \cite{top}. The lighter
quarks and leptons couple to the light Higgs doublet through
higher order nonrenormalizable operators that become effective 
renormalizable operators by the VEVs that are used to cancel the
anomalous $U(1)_A$ $D$--term equation \cite{top}. The novelty
in the construction of ref.\  \cite{fmt}, and in the model of eq.
(\ref{stringmodel}), is that the reduction of the untwisted
Higgs spectrum is obtained by the choice of the boundary
condition basis vectors in eq. (\ref{stringmodel}), whereas
in previous models it was obtained by the choice of
flat directions and analysis of the superpotential \cite{reviewffm}.

The final gauge group of the string model arises
as follows: in the observable sector the NS boundary conditions 
produce gauge group generators for 
\beq
SU(3)_C\times SU(2)_L\times U(1)_C\times U(1)_L\times U(1)_{1,2,3}\times
U(1)_{4,5,6}~~~ .
\label{observablegg}
\eeq
Thus, the $SO(10)$ symmetry is broken to
$SU(3)\times SU(2)_L\times U(1)_C\times U(1)_L$,
where, 
\begin{eqnarray}
U(1)_C& = & {\rm Tr}\, U(3)_C~\Rightarrow~Q_C=
			 \sum_{i=1}^3Q({\bar\psi}^i)~,\label{u1c}\\
U(1)_L& = & {\rm Tr}\, U(2)_L~\Rightarrow~Q_L=
			 \sum_{i=4}^5Q({\bar\psi}^i)~.\label{u1l}
\end{eqnarray}
The flavour $SO(6)^3$ symmetries are broken to $U(1)^{3+n}$ with
$(n=0,\cdots,6)$. The first three, denoted by $U(1)_{j}$ $(j=1,2,3)$, arise 
{}from the world--sheet currents ${\bar\eta}^j{\bar\eta}^{j^*}$.
These three $U(1)$ symmetries are present in all
the three generation free fermionic models which use the NAHE set. 
Additional horizontal $U(1)$ symmetries, denoted by $U(1)_{j}$ 
$(j=4,5,...)$, arise by pairing two real fermions from the sets
$\{{\bar y}^{3,\cdots,6}\}$, 
$\{{\bar y}^{1,2},{\bar\omega}^{5,6}\}$ and
$\{{\bar\omega}^{1,\cdots,4}\}$. 
The final observable gauge group depends on
the number of such pairings. In this model there are the 
pairings ${\bar y}^3{\bar y}^6$, ${\bar y}^1{\bar\omega}^5$
and ${\bar\omega}^2{\bar\omega}^4$, which generate three additional 
$U(1)$ symmetries, denoted by $U(1)_{4,5,6}$. 

It is important to note that the existence of these three additional 
$U(1)$ currents is correlated with the assignment of asymmetric
boundary conditions with respect to the set of internal
world--sheet fermions $\{y,\omega|{\bar y},{\bar\omega}\}^{1,\cdots,6}$,
in the basis vectors that extend the 
NAHE--set, $\{\alpha, \beta,\gamma\}$.
This assignment of asymmetric boundary conditions in the basis
vector that breaks the $SO(10)$ symmetry to $SO(6)\times SO(4)$
results in the projection of the untwisted Higgs colour--triplet fields
and preservation of the  corresponding electroweak--doublet Higgs
representations \cite{ps}.

In the hidden sector, 
which arises from the complex
world--sheet fermions ${\bar\phi}^{1\cdots8}$,
the NS boundary conditions produce the generators of
\beq
SU(2)_{1,2,3,4}\times SU(4)_{H_1}\times U(1)_{H_1}\, .
\label{hiddengg}
\eeq
$U(1)_{H_1}$ 
corresponds to the combinations of the world--sheet charges
\begin{equation}
Q_{H_1}=\sum_{i=5}^8Q({\bar\phi}^i)~.\label{qh1}
\end{equation}

The model contains several additional sectors that may a priori 
produce space--time vector bosons and enhance the gauge symmetry, which include
the sectors
$\zeta\equiv{\bf 1}+b_1+b_2+b_3$ and ${\bf 1}+S+\alpha+\beta+\gamma$.
Additional space--time vector bosons from these sectors would enhance
the gauge symmetry that arise from the space--time vector bosons produced
in the Neveu--Schwarz sector.
However, with the choice of generalised GSO projection coefficients 
given in eq. (\ref{phasesmodel1}) all of the extra gauge bosons from these
sectors are projected out and the four dimensional gauge group is 
given by eqs. (\ref{observablegg}) and (\ref{hiddengg}).

In addition to the graviton, dilaton,
antisymmetric sector and spin--1 gauge bosons, 
the Neveu--Schwarz  sector gives one pair of electroweak Higgs 
doublets $h_3$ and $\bar h_3$; six pairs of $SO(10)$ singlets, which 
are charged with respect to $U(1)_{4,5,6}$;
three singlets of the entire four dimensional gauge group.
A notable difference as compared to models with unreduced
untwisted Higgs spectrum, like the model of ref.\  \cite{eu},
is that the $SO(10)$ singlet fields, which are charged under
$U(1)_{1,2,3}$, are projected out from the massless spectrum.
The three generations are obtained from the sectors $b_1$, $b_2$ and 
$b_3$. The model contain states that are vector--like with respect to the
Standard Model and all non--Abelian group factors, but may be chiral with
respect to the $U(1)$ symmetries that are orthogonal to the $SO(10)$ group. 
The full massless spectrum of the model is detailed in Table 1
at the end of this paper.

As a final note we remark that the boundary conditions with respect to the 
internal world--sheet fermions of the set
$\{y,\omega|{\bar y},{\bar\omega}\}^{1,\cdots,6}$
in the basis vectors $\alpha$, $\beta$ and $\gamma$,
that extend the NAHE--set, are similar to those
in the basis vectors that generate the string 
model of ref.\ \  \cite{eu}, with the replacements
\beqn
\alpha({\bar y}^3{\bar y}^6)& \longleftrightarrow & 
\gamma({\bar y}^3{\bar y}^6) \nonumber\\
\beta({\bar y}^1{\bar\omega}^5)& \longleftrightarrow & 
\gamma({\bar y}^1{\bar\omega}^5) . \label{278substitutions}
\eeqn

The world--sheet 
fermions $\{y,\omega|{\bar y},{\bar\omega}\}^{1,\cdots,6}$
correspond to the compactified dimensions in a corresponding
bosonic formulation.
The substitutions in (\ref{278substitutions})
are augmented with suitable modifications
of the boundary conditions of the world--sheet fermions
$\{{\bar\psi}^{1,\cdots,5},{\bar\eta}^{1,\cdots,3},{\bar\phi}^{1,\cdots,8}\}$,
which correspond to the gauge degrees of freedom.  
The effect of these additional 
modifications is to alter the hidden sector gauge group.
While the substitutions in (\ref{278substitutions})
look innocuous enough, they in fact produce substantial 
changes in the massless spectrum and, as a consequence, in the
physical characteristics of the models. With regard to the
flat directions of the superpotential, the effect of these
changes on the untwisted states will be particularly noted.

\subsection{Third through Fifth Order Superpotential}

The three singlets of the entire four dimensional gauge group are obtained from: 
\beqn
\xi_{1}&=&\chi^{12*}\bar{\omega}^3\bar{\omega}^6|0>~~,\nonumber\\
\xi_{2}&=&\chi^{34*}\bar{\omega}^1\bar{y}^5|0>~~,\nonumber\\
\xi_{3}&=&\chi^{56*}\bar{y}^2\bar{y}^4|0>~~.\nonumber
\eeqn

We show below the cubic through quintic order superpotential terms. 

\noindent Trilinear superpotential:
\beqn
W_3&=&N^c_3 L_3\bar{h}+u^c_3 Q_3\bar{h} + 
		H_4\bar{H}_7 h+\bar{H}_4H_7\bar{h}+\nonumber\\
&+&{\xi_1}(H_1\bar{H}_1+H_8\bar{H}_8+H_9\bar{H}_9)\nonumber\\
&+&{\xi_2}(H_2\bar{H}_2+H_{10}\bar{H}_{10}+H_{11}\bar{H}_{11})\nonumber\\
&+&{\xi_3}(H_3\bar{H}_3+H_4\bar{H}_4+H_5\bar{H}_5+H_6\bar{H}_6+
		H_7\bar{H}_7)\nonumber\\
&+&{\xi_3}(\Phi_1^{\alpha\beta}\bar{\Phi}_1^{\alpha\beta}+
		\Phi_2^{\alpha\beta}\bar{\Phi}_2^{\alpha\beta})\nonumber\\
&+&\Phi_{45}(\bar{\Phi}_{46}\bar{\Phi}'_{56}+\bar{\Phi}'_{46}\bar{\Phi}_{56})+
	\bar{\Phi}_{45}(\Phi_{46}\Phi'_{56}+\Phi'_{46}\Phi_{56})\nonumber\\
&+&\Phi'_{45}(\bar{\Phi}_{46}\Phi_{56}+
	\bar{\Phi}'_{46}\Phi'_{56})+\bar{\Phi}'_{45}(\Phi_{46}\bar{\Phi}_{56}+
	\Phi'_{46}\bar{\Phi}'_{56})\nonumber\\
&+&\Phi'_{45}\left((\Phi_1^{\alpha\beta})^2+(\Phi_2^{\alpha\beta})^2\right)+
	\bar{\Phi}'_{45}\left((\bar{\Phi}_1^{\alpha\beta})^2+
	(\bar{\Phi}_2^{\alpha\beta})^2\right)\nonumber\\
&+&\bar{\Phi}'_{45}H_{12}H_{13}+\Phi_{46}H_{14}H_{15}+
	\bar{\Phi}'_{56}H_{16}H_{17}\nonumber\\
&+&\Phi'_{56}(H_1)^2+\bar{\Phi}'_{56}(\bar{H}_1)^2+
	\bar{\Phi}'_{46}(H_2)^2+ \Phi'_{46}(\bar{H}_2)^2\nonumber\\
&+&\Phi_1^{\alpha\beta}H_9H_{11}+
	\bar{\Phi}_2^{\alpha\beta}(\bar{H}_1\bar{H}_2+\bar{H}_8\bar{H}_{10})+
	 \bar H_1 \bar H_4 H_{10} + H_2 \bar H_4 \bar H_8
\label{w3all}
\eeqn
\noindent Quartic superpotential:
\beqn
W_4&=&  Q_1 u_1 H_4 \bar{H}_5
      + Q_2 u_2 H_4 \bar{H}_6
      + L_1 N^c_1 H_4 \bar{H}_5
      + L_2 N^c_2 H_4 \bar{H}_6
\label{w4all}
\eeqn		
\noindent Quintic superpotential:
\beqn
W_5&=&  
     Q_{1}  H_{3}   L_{1}  \bar{H}_{5}   \xi_{2}
+    Q_{2}  H_{3}   L_{2}  \bar{H}_{6}   \xi_{1}
+    Q_{3}   u^{c}_{3}  \bar{H}_{1}  \bar{H}_{7}  H_{10}
+    Q_{3}   u^{c}_{3}  H_{2}  \bar{H}_{7}  \bar{H}_{8}
\nonumber\\&+&
     d^{c}_{1}   u^{c}_{1}  H_{3}  \bar{H}_{5}   \xi_{2}
+    d^{c}_{1}  H_{3}  H_{3}  \Phi_{46}  V_{2}
+    d^{c}_{2}   u^{c}_{2}  H_{3}  \bar{H}_{6}   \xi_{1}
+    d^{c}_{2}  H_{3}  H_{3}  \bar{\Phi}^{'}_{56}  V_{5}
\nonumber\\&+&
     H_{3}  \bar{H}_{4}  \bar{H}_{1}   \bar{H}_{3}  H_{10}
+    H_{3}  \bar{H}_{4}  H_{2}   \bar{H}_{3}  \bar{H}_{8}
+    H_{3}  \bar{H}_{1}  \bar{H}_{2}   \bar{H}_{3}  \bar{\Phi}^{\alpha\beta}_{2}
+    H_{3}  \bar{H}_{3}  \Phi^{\alpha\beta}_{1}  H_{11}  H_{9}
\nonumber\\&+&
     H_{3}   \bar{H}_{3}  \bar{\Phi}^{\alpha\beta}_{2}  \bar{H}_{8}  \bar{H}_{10}
+    L_{3}  \bar{H}_{1}   N^{c}_3  \bar{H}_{7}  H_{10}
+    L_{3}  H_{2}   N^{c}_3 \bar{H}_{7}  \bar{H}_{8}
+    H_{4}  H_{4} \bar{\Phi}^{'}_{46}  H_{8}  H_{8}
\nonumber\\&+&
     H_{4} H_{4} \Phi_{46}   N^{c}_{1} V_{2}
+    H_{4} H_{4} \bar{\Phi}^{'}_{56}   N^{c}_{2}  V_{5}
+    H_{4} H_{4} \bar{\Phi}^{'}_{56}  \bar{H}_{10}  \bar{H}_{10}
+    H_{4} \bar{H}_{4}  \bar{H}_{4}  \bar{H}_{1}  H_{10}
\nonumber\\&+&
     H_{4} \bar{H}_{4}  \bar{H}_{4}  H_{2}  \bar{H}_{8}
+    H_{4} \bar{H}_{4}  \bar{H}_{1}  \bar{H}_{2}  \bar{\Phi}^{\alpha\beta}_{2}
+    H_{4} \bar{H}_{4}  \Phi^{\alpha\beta}_{1}  H_{11}  H_{9}
+    H_{4} \bar{H}_{4}  \bar{\Phi}^{\alpha\beta}_{2}  \bar{H}_{8}  \bar{H}_{10}
\nonumber\\&+&
     H_{4} H_{1}   \xi_{2}  H03  H_{8}
+    H_{4} H_{2}  \bar{\Phi}^{'}_{56}  \Phi^{\alpha\beta}_{2}  \bar{H}_{10}
+   \bar{H}_{4}  \bar{H}_{4}  \Phi^{'}_{46}  \bar{H}_{8}  \bar{H}_{8}
+   \bar{H}_{4}  \bar{H}_{4}  \Phi^{'}_{56} H_{10}  H_{10}
\nonumber\\&+&
    \bar{H}_{4}  \bar{H}_{1}  \bar{H}_{1}  H_{1}  H_{10}
+   \bar{H}_{4}  \bar{H}_{1}  H_{1}  H_{2}  \bar{H}_{8}
+   \bar{H}_{4}  \bar{H}_{1}  \bar{H}_{2}  H_{2}  H_{10}
+   \bar{H}_{4}  \bar{H}_{1}  H_{7}  \bar{H}_{7}  H_{10}
\nonumber\\&+&
    \bar{H}_{4}  \bar{H}_{1}  H_{6}  \bar{H}_{6}  H_{10}
+   \bar{H}_{4}  \bar{H}_{1}  H_{5}  \bar{H}_{5}  H_{10}
+   \bar{H}_{4}  \bar{H}_{1} \xi_{2} \bar{\Phi}^{\alpha\beta}_{2}  \bar{H}_{8}
+   \bar{H}_{4}  \bar{H}_{1} \Phi^{\alpha\beta}_{1} \bar{\Phi}^{\alpha\beta}_{1} H_{10}
\nonumber\\&+&
   \bar{H}_{4}  \bar{H}_{1} \Phi^{\alpha\beta}_{2} \bar{\Phi}^{\alpha\beta}_{2} H_{10}
+  \bar{H}_{4}  \bar{H}_{1}  H_{11}  H_{10}  \bar{H}_{11}
+   \bar{H}_{4} \bar{H}_{1}  H_{10}  H_{10}  \bar{H}_{10}
+   \bar{H}_{4} \bar{H}_{1}  H_{10}  \bar{H}_{9}  H_{9}
\nonumber\\&+&
    \bar{H}_{4}  \bar{H}_{1}  H_{10}  \bar{H}_{8}  H_{8}
+   \bar{H}_{4}  H_{1}  H_{10}  H_{16}  H_{17}
+   \bar{H}_{4}  \bar{H}_{2}  H_{2}  H_{2}  \bar{H}_{8}
+   \bar{H}_{4}  \bar{H}_{2}  \Phi^{'}_{56} \bar{\Phi}^{\alpha\beta}_{2}  H_{10}
\nonumber\\&+&
    \bar{H}_{4} H_{2} H_{7}  \bar{H}_{7}  \bar{H}_{8}
+   \bar{H}_{4} H_{2} H_{6}  \bar{H}_{6}  \bar{H}_{8}
+   \bar{H}_{4} H_{2} H_{5}  \bar{H}_{5}  \bar{H}_{8}
+   \bar{H}_{4} H_{2} \Phi^{\alpha\beta}_{1} \bar{\Phi}^{\alpha\beta}_{1} 
							\bar{H}_{8}
\nonumber\\&+&
    \bar{H}_{4} H_{2}\Phi^{\alpha\beta}_{2} \bar{\Phi}^{\alpha\beta}_{2}
							\bar{H}_{8}
+   \bar{H}_{4} H_{2} H_{11}  \bar{H}_{8}  \bar{H}_{11}
+   \bar{H}_{4} H_{2} H_{10}  \bar{H}_{8}  \bar{H}_{10}
+   \bar{H}_{4} H_{2}\bar{H}_{9}  \bar{H}_{8}  H_{9}
\nonumber\\&+&
    \bar{H}_{4}  H_{2}  \bar{H}_{8}  \bar{H}_{8}  H_{8}
+   \bar{H}_{1}  \bar{H}_{1}  H_{1}  \bar{H}_{2}  \bar{\Phi}^{\alpha\beta}_{2}
+   \bar{H}_{1}  \bar{H}_{1}  \bar{H}_{2}  \bar{H}_{2}  \Phi^{'}_{45}
+    \bar{H}_{1}  \bar{H}_{1}  \bar{\Phi}^{'}_{46}  \bar{\Phi}^{\alpha\beta}_{1}
                                                   \bar{\Phi}^{\alpha\beta}_{1}
\nonumber\\&+&
   \bar{H}_{1}  \bar{H}_{1}  \bar{\Phi}^{'}_{46}  \bar{\Phi}^{\alpha\beta}_{2}
                                                   \bar{\Phi}^{\alpha\beta}_{2}
+   \bar{H}_{1}  H_{1}  \Phi^{\alpha\beta}_{1}  H_{11}  H_{9}
+   \bar{H}_{1}  H_{1}  \bar{\Phi}^{\alpha\beta}_{2}  \bar{H}_{8}  \bar{H}_{10}
+   \bar{H}_{1}  \bar{H}_{2}  \bar{H}_{2}  H_{2}  \bar{\Phi}^{\alpha\beta}_{2}
\nonumber\\&+&
    \bar{H}_{1}  \bar{H}_{2}  \Phi^{'}_{45}  \bar{H}_{8}  \bar{H}_{10}
+   \bar{H}_{1}  \bar{H}_{2}  H_{7}  \bar{H}_{7}  \bar{\Phi}^{\alpha\beta}_{2}
+   \bar{H}_{1}  \bar{H}_{2}  H_{6}  \bar{H}_{6}  \bar{\Phi}^{\alpha\beta}_{2}
+   \bar{H}_{1}  \bar{H}_{2}  H_{5}  \bar{H}_{5}  \bar{\Phi}^{\alpha\beta}_{2}
\nonumber\\&+&
   \bar{H}_{1}  \bar{H}_{2}  \Phi^{\alpha\beta}_{1}  \bar{\Phi}^{\alpha\beta}_{1}
                                                     \bar{\Phi}^{\alpha\beta}_{2}
+   \bar{H}_{1}  \bar{H}_{2}  \Phi^{\alpha\beta}_{2}  \bar{\Phi}^{\alpha\beta}_{1}
                                                     \bar{\Phi}^{\alpha\beta}_{1}
+   \bar{H}_{1}  \bar{H}_{2}  \Phi^{\alpha\beta}_{2}  \bar{\Phi}^{\alpha\beta}_{2}
                                                     \bar{\Phi}^{\alpha\beta}_{2}
+   \bar{H}_{1}  \bar{H}_{2}  \bar{\Phi}^{\alpha\beta}_{2}  H_{11}  \bar{H}_{11}
\nonumber\\&+&
   \bar{H}_{1}  \bar{H}_{2}  \bar{\Phi}^{\alpha\beta}_{2}  H_{10}  \bar{H}_{10}
+   \bar{H}_{1}  \bar{H}_{2}  \bar{\Phi}^{\alpha\beta}_{2}  \bar{H}_{9}  H_{9}
+   \bar{H}_{1}  \bar{H}_{2}  \bar{\Phi}^{\alpha\beta}_{2}  \bar{H}_{8}  H_{8}
+     H_{1}  H_{1}  H_{2}  H_{2}  \bar{\Phi}^{'}_{45}
\nonumber\\&+&
    H_{1}  H_{1}  \Phi^{'}_{46}  \Phi^{\alpha\beta}_{1}  \Phi^{\alpha\beta}_{1}
+    H_{1}  H_{1}  \Phi^{'}_{46}  \Phi^{\alpha\beta}_{2}  \Phi^{\alpha\beta}_{2}
+    H_{1}  H_{1}  \Phi^{'}_{46}  H_{12}  H_{13}
+     H_{1}  H_{1}  \Phi_{45}  H_{14}  H_{15}
\nonumber\\&+&
    H_{1}  \bar{H}_{2}  \bar{\Phi}^{\alpha\beta}_{2}  H_{16}  H_{17}
+    H_{1}  H_{2}  \bar{\Phi}^{'}_{45}  H_{10}  H_{8}
+    \bar{H}_{2}  \bar{H}_{2}  \Phi^{'}_{45}  H_{16}  H_{17}
+     \bar{H}_{2}  \bar{H}_{2}  \Phi^{'}_{56} \bar{\Phi}^{\alpha\beta}_{1}
                                               \bar{\Phi}^{\alpha\beta}_{1}
\nonumber\\&+&
    \bar{H}_{2}  \bar{H}_{2}  \Phi^{'}_{56} \bar{\Phi}^{\alpha\beta}_{2}
                                               \bar{\Phi}^{\alpha\beta}_{2}
+    \bar{H}_{2}  H_{2}  \Phi^{\alpha\beta}_{1}  H_{11}  H_{9}
+    \bar{H}_{2}  H_{2}  \bar{\Phi}^{\alpha\beta}_{2}  \bar{H}_{8}  \bar{H}_{10}
+   H_{2}  H_{2}\bar{\Phi}^{'}_{56}\Phi^{\alpha\beta}_{1}\Phi^{\alpha\beta}_{1}
\nonumber\\&+&
   H_{2}  H_{2}\bar{\Phi}^{'}_{56}\Phi^{\alpha\beta}_{2}\Phi^{\alpha\beta}_{2}
+  H_{2}  H_{2}  \bar{\Phi}^{'}_{56}  H_{12}  H_{13}
+ \bar{\Phi}^{'}_{46}\bar{\Phi}^{\alpha\beta}_{1} 
		\bar{\Phi}^{\alpha\beta}_{1}H_{16} H_{17}
+ \bar{\Phi}^{'}_{46}\bar{\Phi}^{\alpha\beta}_{2} 
		\bar{\Phi}^{\alpha\beta}_{2}H_{16}  H_{17}
\nonumber\\&+&
    \Phi_{46}   N^{c}_3 V_{9}  \bar{H}_{9}  \bar{H}_{9}
+    \Phi_{46}   N^{c}_3 V_{8}  \bar{H}_{8}  \bar{H}_{8}
+    \Phi^{'}_{45}  N^{c}_{2}  V_{6}  \bar{H}_{9}  \bar{H}_{9}
+    \Phi^{'}_{45}   N^{c}_{2}  V_{5}  \bar{H}_{8}  \bar{H}_{8}
\nonumber\\&+&
    \Phi^{'}_{45}  \bar{H}_{9}  \bar{H}_{9}  \bar{H}_{11}  \bar{H}_{11}
+    \Phi^{'}_{45}  \bar{H}_{8}  \bar{H}_{8}  \bar{H}_{10}  \bar{H}_{10}
+    \bar{\Phi}^{'}_{45}  H_{11}  H_{11}  H_{9}  H_{9}
+    \bar{\Phi}^{'}_{45}  H_{10}  H_{10}  H_{8}  H_{8}
\nonumber\\&+&
     \Phi_{45}   N^{c}_{1} V_{3}  H_{11}  H_{11}
+    \Phi_{45}   N^{c}_{1} V_{2}  H_{10}  H_{10}
+    \Phi_{56}   N^{c}_3 V_{9}  H_{11}  H_{11}
+    \Phi_{56}   N^{c}_3 V_{8}  H_{10}  H_{10}
\nonumber\\&+&
     \Phi_{56}  \bar{\Phi}^{\alpha\beta}_{1}  \bar{\Phi}^{\alpha\beta}_{1}  
		H_{14}  H_{15}
+    \Phi_{56}  \bar{\Phi}^{\alpha\beta}_{2}  \bar{\Phi}^{\alpha\beta}_{2}  
		H_{14}  H_{15}
+    N^{c}_{2}  V_{5}  \bar{\Phi}^{\alpha\beta}_{2}  H_{10}  \bar{H}_{8}
+    N^{c}_{2}  \bar{\Phi}^{\alpha\beta}_{2}  H_{11}  \bar{H}_{8}  V_{12}
\nonumber\\&+&
     H_{7}  \bar{H}_{7}  \Phi^{\alpha\beta}_{1}  H_{11}  H_{9}
+    H_{7}  \bar{H}_{7}  \bar{\Phi}^{\alpha\beta}_{2}  \bar{H}_{8}  \bar{H}_{10}
+    H_{6}  \bar{H}_{6}  \Phi^{\alpha\beta}_{1}  H_{11}  H_{9}
+    H_{6}  \bar{H}_{6}  \bar{\Phi}^{\alpha\beta}_{2}  \bar{H}_{8}  \bar{H}_{10}
\nonumber\\&+&
     H_{5}  \bar{H}_{5}  \Phi^{\alpha\beta}_{1}  H_{11}  H_{9}
+    H_{5}  \bar{H}_{5}  \bar{\Phi}^{\alpha\beta}_{2}  \bar{H}_{8}  \bar{H}_{10}
+   \Phi^{\alpha\beta}_{1}  \Phi^{\alpha\beta}_{1}  \bar{\Phi}^{\alpha\beta}_{1}  
		H_{11}  H_{9}
+   \Phi^{\alpha\beta}_{1}  \Phi^{\alpha\beta}_{2}  \bar{\Phi}^{\alpha\beta}_{2}  
		H_{11}  H_{9}
\nonumber\\&+&
    \Phi^{\alpha\beta}_{1}  \bar{\Phi}^{\alpha\beta}_{1}  
		\bar{\Phi}^{\alpha\beta}_{2} \bar{H}_{8}  \bar{H}_{10}
+   \Phi^{\alpha\beta}_{1}  H_{11}  H_{11}  \bar{H}_{11}  H_{9}
+    \Phi^{\alpha\beta}_{1}  H_{11}  H_{10}  \bar{H}_{10}  H_{9}
+     \Phi^{\alpha\beta}_{1}  H_{11}  \bar{H}_{9}  H_{9}  H_{9}
\nonumber\\&+&
    \Phi^{\alpha\beta}_{1}  H_{11}  \bar{H}_{8}  H_{9}  H_{8}
+   \Phi^{\alpha\beta}_{2}  \Phi^{\alpha\beta}_{2}  
		\bar{\Phi}^{\alpha\beta}_{1}  H_{11}  H_{9}
+   \Phi^{\alpha\beta}_{2}  \bar{\Phi}^{\alpha\beta}_{1}  
		\bar{\Phi}^{\alpha\beta}_{1} \bar{H}_{8}  \bar{H}_{10}
+   \Phi^{\alpha\beta}_{2}  \bar{\Phi}^{\alpha\beta}_{2}  
		\bar{\Phi}^{\alpha\beta}_{2} \bar{H}_{8}  \bar{H}_{10}
\nonumber\\&+&
     \bar{\Phi}^{\alpha\beta}_{1}  H_{11}  H_{12}  H_{9}  H_{13}
+    \bar{\Phi}^{\alpha\beta}_{2}  H_{11}  \bar{H}_{8}  \bar{H}_{11}  \bar{H}_{10}
+    \bar{\Phi}^{\alpha\beta}_{2}  H_{10}  \bar{H}_{8}  \bar{H}_{10}  \bar{H}_{10}
+    \bar{\Phi}^{\alpha\beta}_{2}  \bar{H}_{9}  \bar{H}_{8}  \bar{H}_{10}  H_{9}
\nonumber\\&+&
     \bar{\Phi}^{\alpha\beta}_{2}  \bar{H}_{8}  \bar{H}_{8}  \bar{H}_{10}  H_{8}
\label{w5all}\\
\nonumber
\eeqn		

\section{Flat directions}\label{flatdir}

The model possesses nine local $U(1)$ symmetries, 
eight in the observable part
and one in the hidden part. Six of these are anomalous:
\beq
{\rm Tr}{U_1}= {\rm Tr}{U_2}= - {\rm Tr}{U_3}=  
2 {\rm Tr}{U_4}= -2 {\rm Tr}{U_5}= 2 {\rm Tr}{U_6}= -24.
\label{tru1s}
\eeq
$U(1)_L$ and $U(1)_C$ of the $SO(10)$ subgroup are anomaly free.  
Consequently, the weak hypercharge and the orthogonal combination,
$U(1)_{Z^{\prime}}$, are  anomaly free.
The hidden sector $U(1)_{H_1}$ is also anomaly free.\

Of the six anomalous $U(1)$s, five can be rotated by
an orthogonal transformation to become anomaly free. 
The unique combination that remains anomalous is: 
${U_A}=k\sum_j [{{\rm Tr} {U(1)_j}}]U(1)_j$, 
where $j$ runs over all the anomalous $U(1)$s and $k$ is
a normalisation constant. 
For convenience, we take $k={1\over{12}}$ and therefore 
the anomalous combination is given by:
\beq
U_A=-2U_1-2U_2+2U_3-U_4+U_5-U_6,{\hskip .5cm}{\rm Tr}Q_A=180.\label{u1a}
\eeq
We note that the anomalous $U(1)_A$ combination in (\ref{u1a}) 
is similar to that of the model of ref.\  \cite{eu}, aside from 
the changes of the sign of the traces over $U(1)_{1,2}$ and $U(1)_5$.

The five rotated non-anomalous orthogonal combinations are not unique,
with different
choices related by orthogonal transformations. One choice is given by:
\beqn
 {U^\prime}_1 & = & U_1-U_2{\hskip .5cm},{\hskip .5cm}
 {U^\prime}_2  ~=~  U_1+U_2+2U_3,\label{u1pu2p}\\
 {U^\prime}_3 & = & U_4+U_5{\hskip .5cm},{\hskip .5cm}
 {U^\prime}_4  ~=~  U_4-U_5-2U_6,\label{u3pu4p}\\
 {U^\prime}_5 & = & U_1+U_2-U_3-2U_4+2U_5-2U_6.\label{u5p}
\eeqn
Thus, after this rotation there are a total of eight $U(1)$s
free from gauge and gravitational anomalies.

A basis set of (norm-squares of) VEVs of scalar fields
satisfying the non-anomalous 
$D$-flatness constraints (\ref{dxi}) can be created en masse 
\cite{systematic1, systematic2, mshsm}. The basis directions can have
positive, negative, or zero anomalous charge. 
In the maximally orthogonal basis used in the singular value decomposition 
approach of \cite{systematic2, mshsm}, 
each basis direction is uniquely identified with a particular VEV. That is, 
although each basis direction generally contains many VEVs, each basis direction 
contains at least one particular VEV that only appears in it.

A physical $D$-flat direction $D_{\rm phys}$, with anomalous charge of sign 
opposite that of the FI term  $\xi$, is formed from linear combinations of the 
basis directions, 
\beqn
D_{\rm phys} = \sum_{\rm i = 1}^{\#~{\rm basis~dirs.}} a_i D_i,
\label{phydir}
\eeqn
where the integer coefficients $a_i$ are normalised to have no non-trivial 
common factor.

In our notation, a physical flat direction (\ref{phydir}) may have a negative 
norm-square for a vector-like field. This denotes that it is the oppositely 
charged vector-partner field that acquires the VEV, rather than the field. 
Basis directions themselves may have vector-like partner directions if all 
associated fields are vector-like. On the other hand, if in particular, the 
field generating the VEV uniquely associated with a basis direction does not 
have a vector-like partner, that basis direction cannot have a vector-like 
partner direction. 

In pursuit of physical all-order flat directions for this model, we first 
examined directions formed solely from the VEVs of non-Abelian singlet fields. An 
associated maximally orthogonal basis set, denoted by
$\{{\cal{D}}^{'}_{i=1\ {\rm to}\ 13}\}$, containing only non-Abelian singlet 
VEVs is shown in Table 2.a. The respective unique VEV fields of these basis 
directions are identified in Table 2.b. Examination of Tables 2.a and 2.b 
reveals that no physical $D$-flat directions can be formed solely from VEVs
of non-Abelian singlet fields.
Since the FI term $\xi$ (\ref{dxi}) is positive for this model, with ${\rm 
Tr}Q_A=180$, a physical flat direction must carry a negative anomalous charge. 
However, of the 13 singlet $D$-flat basis directions, three carry anomalous 
charge of $+15$, $+30$, $+30$ while the remaining ten do not carry anomalous charge. Further,
the unique VEVed fields for the 3 basis directions with positive anomalous charge 
do not have corresponding vector-like partner fields. Hence, there are no 
vector-like paired basis directions with negative anomalous charge. Thus, Tables 
2.a and 2.b imply that one or more fields carrying non-Abelian charges must also 
acquire VEVs in physical $D$-flat directions. This result is, in itself, not 
necessarily unexpected, as non-Abelian VEVs have been required for physical 
(all-order) flat directions in other quasi-realistic free fermionic heterotic 
models in the past, for example \cite{cfs}. 

Thus, we expanded our flat direction search to include VEVs of both non-Abelian
singlet fields and non-Abelian charged fields. Our chosen set of 50 maximally
orthogonal $D$-flat basis directions for both non-Abelian singlet VEVs and
non-Abelian charged VEVs, denoted by $\{{\cal{D}}_{i=1\ {\rm to}\ 50}\}$, is
presented in Table 3.a. The respective unique field VEVs identified with these
basis directions are given in Table 3.b. In this enlarged basis the anomalous 
charges are given in units of ($\frac{Q^{(A)}}{15}$) and the directions 
containing only singlet VEVs are rotations of those in Table 2.a.

Nine of the 50 directions, denoted $D_{i=1,...,9}$, carry one or two units of
negative anomalous charge. Twenty basis directions, denoted $D_{10}$ through
$D_{29}$, carry no anomalous charge. Twenty-one basis directions, denoted
$D_{30}$ through $D_{50}$, carry one or two units of positive anomalous charge.  
All basis directions possessing negative anomalous charge contain $SU(3)_C\otimes
SU(2)_L$ charges or hidden sector $SU(4)\otimes \prod^{4}_{j=1} SU(2)_j$ charges.
(Thus, this basis set also reveals that anomaly cancellation will necessarily
break one or more non-Abelian local symmetries.) All of the $\Phi$ fields, the
$H_{1~{\rm to}~11}$ fields and $h$ have vector-like pairs. Thus, physical flat
directions can have negative components for any of these. A subset of these
fields, specifically $\Phi_{46}$, $\Phi^{'}_{45}$, $\bar{\Phi}^{'}_{56}$,
and $H_{4,5,6,7}$,
has VEVs appearing in multiple basis directions. The only non-vector-like field
with a VEV that appears in multiple directions is $e^c_{3}$.
 
$D_{10}$ through $D_{17}$ and $D_{22}$ are composed solely of varying
combinations of the vector-like fields. Hence, all of these basis directions have
corresponding vector-like partner basis directions, $\bar{D}_{i} \equiv -D_{i}$, 
for which the VEV of each field is replaced by the VEV of the vector-like partner
field. Thus, in a physical flat direction (\ref{phydir}), each of the respective
integer coefficients $a_{10}$ through $a_{17}$ and $a_{22}$, may be
negative, positive, or zero.

Note that $D_{7}$, $D_{8}$, $D_{9}$ and $D_{20}$ are vector-like except for their 
$e^c_{3}$ components. Thus, each of $a_{7}$, $a_{8}$, $a_9$ and $a_{20}$ may be 
negative, positive, or zero in a physical $D$-flat direction, so long as the net
norm-square VEV of $e^c_{3}$ is non-negative.\footnote{Note that non-vector-like
fields, such as $e^c_3$, that appear in multiple directions with some basis directions
having positive and some having negative norm-square components,
are common in this process.
Further, some models explored in the past have had (at least) one basis direction with two
(or more) field VEVs unique to it and with norm-square VEVs with differing signs.
This latter type of basis direction can never appear in a physical direction and, hence,
implies that the fields unique to it can never appear in a $D$-flat
direction. (If all of the
norm-squares of the fields unique to a basis direction were initially negative, then these
signs, along with those of the norm-squares of any vector-like field VEVs in that basis
direction, could all be changed together to allow the basis direction
to appear in a physical
direction.)} The remaining basis directions
contain at least one unique non-vector-like field VEV. Thus, in a physical flat
direction, the coefficients of the remaining basis directions must be 
non-negative.

What does this mean for a physical $D$-flat direction formed as a linear
combination of the basis directions? For a physical flat direction there are,
thus, two specific constraints on the $a_i$ coefficients and one general set of
non-negative norm-square constraints on a subset of the $a_i$. First, negative
anomalous charge for a flat direction requires
\beqn
-2 \sum_{i=1}^2 a_i     -   \sum_{i=3}^{9} a_i 
+  \sum_{i=30}^{44} a_i + 2 \sum_{i=45}^{50} a_i < 0.
\label{physdc1} 
\eeqn 
Second, a non-negative norm-square VEV for $e^c_3$ requires 
\beqn
&&-6 \sum_{i=1}^{2} a_i -3 a_3 - 6\sum_{i=4}^{6} a_i -2 a_7 -6 \sum_{i=8}^{9} a_i 
-2 \sum_{i=18}^{19} a_i - a_{20} + a_{21}\nolabel\\ 
&&-2 \sum_{i=23}^{24} a_i - a_{25} -2 a_{26} +2 a_{27} - 2 a_{28} +2 a_{29}
+6 a_{30} +6 a_{32}+ a_{38} \nolabel\\
&&+ 3\sum_{i=39}^{40} a_i
+6 a_{42} + 6 \sum_{i=45}^{47} a_i
+2\sum_{i=48}^{49} a_i + 6 a_{50}\geq 0.
\label{physdc2} 
\eeqn 
Last, for the set of 
non-vector-like fields that are each identified with a respective unique
$D$-flat direction, the general set of non-negative norm-square VEV constraints 
is \beqn
a_i~\geq 0~{\rm for}~i=1~{\rm to}~6,~18,~19,~21,~23~{\rm to}~50.
\label{physdc3}
\eeqn

At low orders, each individual superpotential term also induces several
stringent $F$-term  constraints on the $a_i$ coefficients of physical flat
directions. As stated prior, 
the set of constraints from superpotential terms with only singlet
fields translate into the requirement that two or more singlet fields in a given
superpotential term cannot take on VEVs. For the model under investigation,
constraints from third order superpotential terms are especially severe. For 
this model, all six $\Phi$ singlet fields and their vector-like partners appear in
third order superpotential terms (specifically, the sixth and seventh lines) of 
(\ref{w3all}). Stringent $F$-flatness from these terms forbids at least 8 of the
12 singlet fields from acquiring VEVs.

For example, when solely third order stringent $F$-flatness constraints are
applied to the six pairs of $\Phi$ vector-like singlets (and no $F$-flatness
constraints are applied to the non-Abelian states), there are just nine solution
classes that allow the
maximum of 4 singlet VEVs. (Flat directions in any of these nine classes are
defined by their respective non-Abelian VEVs.)

For three of these nine singlet third order flatness classes, the VEVs are of two fields
and their respective vector-like partners: either,
\beqn
<{\Phi}_{45}>,\, <{\Phi}'_{45}>,\, <\bar{\Phi}_{45}>,\,
<\bar{\Phi}'_{45}> &\ne 0,&\, {\rm or}
\label{fflat3sa}\\
<\bar{\Phi}_{46}>,\, <\bar{\Phi}'_{46}>,\, 
<{\Phi}_{46}>,\, <{\Phi}'_{46}> &\ne 0,&\, 
{\rm or}
\label{fflat3sb}\\
<\bar{\Phi}'_{56}>,\, <\bar{\Phi}_{56}>,\, 
<{\Phi}'_{56}>,\, <{\Phi}_{56}> &\ne 0.&
\label{fflat3sc}
\eeqn
Higher order stringent flatness constraints can further reduce the allowed number
of singlet VEVs of each of these solutions. Further, a component of a $D$-flat
basis direction in Table 3.a only specifies the difference between the 
norm-squares of the VEV of a given field and of the given vector-like partner
field (if it exists). Completely chargeless VEVs solely involving a field 
$\Phi_i$ and its vector-like partner $\bar{\Phi}_i$ such that $|<\Phi_i>|^2 = 
|<\bar{\Phi}_i>|^2$ can always be added to a physical $D$-flat direction. 
However, it is preferable for higher order $F$-flatness to impose that a field
and its vector-partner do not simultaneously acquire VEVs. Hence, these three
solutions effectively allow only two unique singlet fields to acquire VEVs.

The next three classes of singlet solutions do allow up to four distinct
singlet fields to
acquire VEVs: either,
\beqn
&&<{\Phi}_{45}>,\, <{\Phi}'_{45}>,\, <{\Phi}_{46}>,\, <{\Phi}'_{46}> \ne 0,
{\rm or},
\label{fflat3sd}\\
&&<{\Phi}_{45}>,\, <{\Phi}'_{56}>,\, 
<{\Phi}_{56}>\, <\bar{\Phi}'_{45}>   \ne 0,\, 
{\rm or},
\label{fflat3se}\\
&&<\bar{\Phi}_{46}>,\, <\bar{\Phi}_{56}>,\, <{\Phi}'_{56}>,\, <{\Phi}'_{46}>
\ne 0.
\label{fflat3sf}
\eeqn

For the three remaining solution classes, the fields in (\ref{fflat3sd}),
(\ref{fflat3se}) 
and (\ref{fflat3sf}), are respectively replaced with their vector-like partner
fields. For any of these nine stringent $F$-flat choices, no other $\Phi$ singlet 
fields can acquire VEVs.

Any of the constraints on allowed and disallowed VEVs, such as the above, can be
re-expressed in terms of constraints on the $a_i$ coefficients specifying the
basis directions contributions to a physical $D$-flat direction. For example,
setting $<\Phi_{46}>=0$ would require
\beqn
&&4 a_1 + a_2  +2 \sum_{i=3}^{4} a_i +8 a_5 +2 a_6 + 
a_7 - a_8 - a_9 + a_{10} + a_{16}  \nolabel\\ 
&&- a_{18} +2 a_{19} + a_{20} - a_{21} + a_{23}-2 a_{27} - a_{28} + 
a_{29} + 4 a_{30} \nolabel\\
&&+ a_{31}  - a_{32} +\sum_{i=33}^{35} a_i -2 \sum_{i=36}^{37} a_i -
2 a_{39} + a_{40} 
-2 \sum_{i=41}^{43} a_i + a_{44}\nolabel\\
&& -4 a_{45} + 2 a_{46} - a_{47} -3 a_{49} - a_{50} = 0 ~ ~ .
\label{examp1} 
\eeqn

To systematically investigate physical $D$-flat directions with non-Abelian VEVs,
over a course of eight months we generated and examined physical $D$-flat
directions composed of from 1 to 6 basis directions. Under the assumption that
all VEVs of physical flat directions are nearly of the same order of magnitude,
we allowed coefficients of 0 to 20 for the non-vector-like basis directions and
coefficients of -20 to 20 for the vector-like basis directions.

To be classified as a physical $D$-flat direction, a linear combinations of basis
directions needed to obey (\ref{physdc1}-\ref{physdc3}) and was, of course, 
also required to have non-Abelian $D$-flatness. (The general process by
which we enforced non-Abelian $D$-flatness followed that presented in
\cite{systematic2,mshsm}.) Each resulting physical $D$-flat direction was then
tested for stringent $F$-flatness from all third order through fifth order
superpotential terms and additionally for some key sixth order superpotential
terms.\footnote{While only the third through fifth order superpotential is given
in section 4.1, we have generated the complete superpotential to eighth order
and can generate it to any required order.}
 
Following the SVD method discussed earlier in section 3 and described in
\cite{cfn,cfnw,Cleaver:2007ek},
we had planned to then test for possible all-order stringent $F$-flatness, the 
subset of physical $D$-flat directions that had proven stringently $F$-flat to
at least fifth or sixth order. Based on all of the prior models we had
investigated, we had expected to find around four to six physical $D$-flat
directions that were, in fact, stringently $F$-flat to all finite order.
However, in contrast we discovered that no physical $D$-flat directions that we
had generated even kept stringent
$F$-flatness through sixth order. So there were no
physical $D$-flat directions to examine for all-order testing! For this model,
with its reduced 
set of singlet fields from the untwisted sector, not even self-cancellation of 
non-Abelian terms could provide stringent $F$-flatness through sixth order for 
any of these physical $D$-flat directions.

We will continue for several years a search for $F$-flatness past sixth order
for physical $D$-flat directions in this model that are comprised of seven or
more basis directions. After January 2008, the search will be conducted on a
128-node computer system of quad-processors to quicken the pace of the
investigation. However, a continued null result is likely: since each of our
basis directions contains a unique field VEV, increasing the number of non-zero
$a_i$ coefficients linearly increases the minimum number of unique field VEVs.
With each increase in number of basis directions composing a physical $D$-flat
direction, the probability of obtaining stringent $F$-flatness much beyond sixth
order further decreases.

\section{Conclusions}\label{conclude}

The string models in the free fermionic formulation gave rise to a large 
class of quasi--realistic string models, including three generation models
that produce solely the MSSM spectrum in the observable 
Standard Model charged sector of the effective low energy field theory.
As such the free fermionic models provides the arena to study how string
theory may be related to the observed particle data. In turn,
the properties of the models that make them attractive from the point of 
view of the phenomenological data may be instrumental to 
uncover unexpected properties of string theory.

In this paper we stumbled upon such a possible novel string feature
with regard to supersymmetry breaking. A general expectation from 
supersymmetric quantum field theories is that if supersymmetry 
is unbroken at the lowest order in perturbation theory, then it cannot
be broken at higher perturbative orders in quantum perturbation theory.
The model that we presented in this paper opens up the possibility that 
string theory may afford other options.

The model is a quasi--realistic three generation string model in the
free fermionic
formulation that shares many of the characteristics of previous
quasi--realistic 
free fermionic models. It contains three chiral generations charged
under the standard--like model subgroup of the underlying $SO(10)$ symmetry 
of the NAHE--set. The weak--hypercharge possesses the canonical GUT embedding
and the model predicts $\sin^2\theta_W=3/8$ at the unification scale.
The Higgs spectrum contains one untwisted electroweak doublet 
pair that couples at the cubic level of the superpotential to the top quark.
Like numerous other quasi--realistic string models, the model
contains an anomalous 
$U(1)$ symmetry which generates a Fayet--Iliopoulos $D$--term.
The Fayet--Iliopoulos
term would, in general, break supersymmetry, unless there is a direction in the 
scalar potential which is both $D$-- and $F$--flat. If such a direction exists
it will acquire a VEV, cancel the Fayet--Iliopoulos $D$--term and 
restore supersymmetry. The general expectation, due to the corresponding
results in supersymmetric point quantum field theories, is that a
supersymmetric vacuum does exist due to the fact that
string spectrum is Bose--Fermi
degenerate and possesses $N=1$ supersymmetry. 

Indeed, in all the previously studied
quasi--realistic free fermionic models such supersymmetric flat directions
were found. Moreover, all previous models yielded the so called stringent 
flat directions that can be shown to be exact, {\it i.e.} flat to all
orders of nonrenormalizable terms. The distinct feature of the
model discussed in this paper is that it does not admit such stringent flat
directions. In this model no physical $D$--flat direction that we generated
kept $F$--flatness through sixth order.
We speculate that only stringent flat directions can be flat to all orders
of nonrenormalizable terms. If it is validated, this would indicate that 
this model, therefore, appears to have no $D$--flat directions that can be
proven to be $F$--flat to all orders, other than by order by order analysis.

As we discussed this outcome may be a general result of the
assignment of boundary conditions to the internal world sheet fermions,
which results in the projection of two of the untwisted Higgs pairs.
If a non--vanishing $F$--term does exist, the implication would be that supersymmetry
remains unbroken at finite order. The Fayet--Iliopoulos term that
breaks supersymmetry is generated at the one--loop level in the
perturbative string expansion. On the other hand the string spectrum
is Bose--Fermi degenerate and possesses $N=1$ space--time supersymmetry
at the classical level. This would suggest that,
contrary to the expectation from
supersymmetric point quantum field theories,
perturbative supersymmetry breaking
in string theory may ensue. 
Furthermore, the modular invariant
one--loop partition function vanishes and, hence, the cosmological 
constant vanishes at the one loop level as well. The model presented 
here may therefore represent an example of a
quasi--realistic string vacuum with 
vanishing one--loop cosmological constant and perturbatively
broken supersymmetry. 
Furthermore, the asymmetric boundary conditions of the string model
given in eq. (\ref{stringmodel}) project all the geometrical moduli
of the underlying $Z_2\times Z_2$ orbifold \cite{modulifixing,foc}.
Absence of supersymmetric flat solutions would imply that the supersymmetric
moduli are fixed as well in this model. Examining the hidden
sector gauge group of the model, given in eq. (\ref{hiddengg}), 
we note that it contains $SU(2)^4$ and satisfies the conditions for the
dilaton race--track stabilisation mechanism \cite{racetrack}. 
It remains without saying that many issues
still need to be further explored
and understood to ascertain the claims of this paper.
Nevertheless, the 
free fermionic models continue to generate intriguing and exciting results
that may at the end prove to be relevant to the elucidation of the connection
between string theory and the real world.

\bigskip
\medskip
\leftline{\large\bf Acknowledgements}
\medskip

AEF would like to thank the CERN theory division for hospitality. 
This work was supported by the STFC, by the University of Liverpool 
and by Baylor University.


\def\MPLA#1#2#3{{\it Mod.\ Phys.\ Lett.}\/ {\bf A#1} (#2) #3}
\def\IJMP#1#2#3{{\it Int.\ J.\ Mod.\ Phys.}\/ {\bf A#1} (#2) #3}
\def\IJMPA#1#2#3{{\it Int.\ J.\ Mod.\ Phys.}\/ {\bf A#1} (#2) #3}



\bigskip
\medskip

\bibliographystyle{unsrt}


\textwidth=7,5 in
\oddsidemargin=-10mm
\topmargin=10mm
\renewcommand{\baselinestretch}{1.3}
\begin{table}
\begin{eqnarray*}
\begin{tabular}{|c|c|c|rrrrrrrr|c|c|}
\hline
  $F$ & SEC & $SU(3)\times$&$Q_{C}$ & $Q_L$ & $Q_1$ &
   $Q_2$ & $Q_3$ & $Q_{4}$ & $Q_{5}$ & $Q_6$ &
   $SU(2)_{1,..,4} $ & $Q_{H_1}$  \\
   $$ & $$ & $SU(2)$ & $$ & $$ & $$ & $$ & $$ & $$ & $$ & $$ & $\times 
							SU(4)_{H_1}$ & $$  \\
\hline
   $L_1$ & $b_1$      & $(1,2)$ & $-{3\over2}$ & $0$ &
   $-{1\over2}$ & $0$ & $0$ & $1\over2$ & $0$ & $0$ &
   $(1,1,1,1,1)$ & $0$  \\
   $Q_1$ &            & $(3,2)$ & $ {1\over2}$ & $0$ &
   $-{1\over2}$ & $0$ & $0$ & $-{1\over2}$ & $0$ & $0$ &
   $(1,1,1,1,1)$ & $0$  \\
   ${d}_{1}^c$ &            & $({\bar 3},1)$ & $-{1\over2}$ & $1$ &
   $-{1\over2}$ & $0$ & $0$ & $-{1\over2}$ & $0$ & $0$ &
   $(1,1,1,1,1)$ & $0$  \\
   ${N}_{1}^c $&            & $(1,1)$ & ${3\over2}$ & $-1$ &
   $-{1\over 2}$ & $0$ & $0$ & $-{1\over2}$ & $0$ & $0$ &
   $(1,1,1,1,1)$ & $0$   \\
   ${u}_{1}^c$ &            & $({\bar 3},1)$ & $-{1\over2}$ & $-1$ &
   $-{1\over2}$ & $0$ & $0$ & $1\over2$ & $0$ & $0$ &
   $(1,1,1,1,1)$ & $0$   \\
   ${e}_{1}^c$ &            & $(1,1)$ & ${3\over2}$ & $1$ &
   $-{1\over 2}$ & $0$ & $0$ & $1\over2$ & $0$ & $0$ &
   $(1,1,1,1,1)$ & $0$   \\
\hline
   $L_2$ & $b_2$      & $(1,2)$ & $-{3\over2}$ & $0$ &
   $0$ & $-{1\over2}$ & $0$ & $0$ & $-{1\over2}$ & $0$ &
   $(1,1,1,1,1)$ & $0$   \\
   $Q_2$ &            & $(3,2)$ & $ {1\over2}$ & $0$ &
   $0$ & $-{1\over2}$ & $0$ & $0$ & ${1\over2}$ & $0$ &
   $(1,1,1,1,1)$ & $0$   \\
   $d_{2}^c$ &            & $({\bar 3},1)$ & $-{1\over2}$ & $1$ &
   $0$ & $-{1\over2}$ & $0$ & $0$ & ${1\over2}$ & $0$ &
   $(1,1,1,1,1)$ & $0$   \\
   ${N}_{2}^c$ &            & $(1,1)$ & ${3\over2}$ & $-1$ &
   $0$ & $-{1\over 2}$ & $0$ & $0$ & ${1\over2}$ & $0$ &
   $(1,1,1,1,1)$ & $0$   \\
   $u_{2}^c$ &            & $({\bar 3},1)$ & $-{1\over2}$ & $-1$ &
   $0$ & $-{1\over2}$ & $0$ & $0$ & $-{1\over2}$ & $0$ &
   $(1,1,1,1,1)$ & $0$   \\
   ${e}_{2}^c$ &            & $(1,1)$ & ${3\over2}$ & $1$ &
   $0$ & $-{1\over 2}$ & $0$ & $0$ & $-{1\over2}$ & $0$ &
   $(1,1,1,1,1)$ & $0$   \\
\hline
   $L_3$ & $b_3$      & $(1,2)$ & $-{3\over2}$ & $0$ &
   $0$ & $0$ & ${1\over2}$ & $0$ & $0$ & $1\over2$ &
   $(1,1,1,1,1)$ & $0$   \\
   $Q_3$ &            & $(3,2)$ & $ {1\over2}$ & $0$ &
   $0$ & $0$ & ${1\over2}$ & $0$ & $0$ & $-{1\over2}$ &
   $(1,1,1,1,1)$ & $0$   \\
   $d_{3}^c$ &            & $({\bar 3},1)$ & $-{1\over2}$ & $1$ &
   $0$ & $0$ & ${1\over2}$ & $0$ & $0$ & $-{1\over2}$ &
   $(1,1,1,1,1)$ & $0$   \\
   ${N}_{3}^c$ &            & $(1,1)$ & ${3\over2}$ & $-1$ &
   $0$ & $0$ & ${1\over 2}$ & $0$ & $0$ & $-{1\over2}$ &
   $(1,1,1,1,1)$ & $0$   \\
   $u_{3}^c$ &            & $({\bar 3},1)$ & $-{1\over2}$ & $-1$ &
   $0$ & $0$ & ${1\over2}$ & $0$ & $0$ & $1\over2$ &
   $(1,1,1,1,1)$ & $0$   \\
   ${e}_{3}^c$ &            & $(1,1)$ & ${3\over2}$ & $1$ &
   $0$ & $0$ & ${1\over 2}$ & $0$ & $0$ & $1\over2$ &
   $(1,1,1,1,1)$ & $0$   \\
\hline
   $h~~$ & ${\rm NS}$ & $(1,2)$ & $0$ & $-1$ &
   $0$ & $0$ & $1$ & $0$ & $0$ & $0$ &
   $(1,1,1,1,1)$  & $0$ \\
   $\bar h~~$ & $$ & $(1,2)$ & $0$ & $1$ &
   $0$ & $0$ & $-1$ & $0$ & $0$ & $0$ &
   $(1,1,1,1,1)$  & $0$  \\
   $\Phi_{56}$ &                       & $(1,1)$ & $0$ & $0$ &
   $0$ & $0$ & $0$ & $0$ & $1$ & $1$ &
   $(1,1,1,1,1)$  & $0$ \\
 $\bar\Phi_{56}$ &                       & $(1,1)$ & $0$ & $0$ &
   $0$ & $0$ & $0$ & $0$ & $-1$ & $-1$ &
   $(1,1,1,1,1)$   & $0$\\
   $\Phi_{56}'$ &                       & $(1,1)$ & $0$ & $0$ &
   $0$ & $0$ & $0$ & $0$ & $1$ & $-1$ &
   $(1,1,1,1,1)$  & $0$ \\
   $\bar\Phi_{56}'$ &                       & $(1,1)$ & $0$ & $0$ &
   $0$ & $0$ & $0$ & $0$ & $-1$ & $1$ &
   $(1,1,1,1,1)$  & $0$ \\
 $\bar\Phi_{46}$ &                       & $(1,1)$ & $0$ & $0$ &
   $0$ & $0$ & $0$ & $-1$ & $0$ & $-1$ &
   $(1,1,1,1,1)$   & $0$\\
   $\Phi_{46}'$ &                       & $(1,1)$ & $0$ & $0$ &
   $0$ & $0$ & $0$ & $1$ & $0$ & $-1$ &
   $(1,1,1,1,1)$   & $0$\\
   $\bar\Phi_{46}'$ &                       & $(1,1)$ & $0$ & $0$ &
   $0$ & $0$ & $0$ & $-1$ & $0$ & $1$ &
   $(1,1,1,1,1)$   & $0$\\
   $\Phi_{46}$ &                       & $(1,1)$ & $0$ & $0$ &
   $0$ & $0$ & $0$ & $1$ & $0$ & $1$ &
   $(1,1,1,1,1)$   & $0$\\
   $\xi_{1,2,3}~~$ &                 & $(1,1)$ & $0$ & $0$ &
   $0$ & $0$ & $0$ & $0$ & $0$ & $0$ &
   $(1,1,1,1,1)$   & $0$\\
   \hline
\end{tabular}
\label{matter1}
\end{eqnarray*}
Table 1. States with charges.
\end{table}

\renewcommand{\baselinestretch}{1.3}
\begin{table}
\begin{eqnarray*}
\begin{tabular}{|c|c|c|rrrrrrrr|c|c|}
\hline
  $F$ & SEC & $SU(3)\times$&$Q_{C}$ & $Q_L$ & $Q_1$ &
   $Q_2$ & $Q_3$ & $Q_{4}$ & $Q_{5}$ & $Q_6$ &
   $SU(2)_{1,..,4} $ & $Q_{H_1}$  \\
   $$ & $$ & $SU(2)$ & $$ & $$ & $$ & $$ & $$ & $$ & $$ & $$ & $\times
							SU(4)_{H_1}$  & $$  \\
\hline
  $\Phi_{45}$ & NS                      & $(1,1)$ & $0$ & $0$ &
   $0$ & $0$ & $0$ & $1$ & $1$ & $0$ &
   $(1,1,1,1,1)$   & $~~0$\\
 $\bar\Phi_{45}$ &                       & $(1,1)$ & $0$ & $0$ &
   $0$ & $0$ & $0$ & $-1$ & $-1$ & $0$ &
   $(1,1,1,1,1)$   &$~~0$\\
   $\Phi_{45}'$ &                       & $(1,1)$ & $0$ & $0$ &
   $0$ & $0$ & $0$ & $1$ & $-1$ & $0$ &
   $(1,1,1,1,1)$   &$~~0$\\
   $\bar\Phi_{45}'$ &                       & $(1,1)$ & $0$ & $0$ &
   $0$ & $0$ & $0$ & $-1$ & $1$ & $0$ &
   $(1,1,1,1,1)$   &$~~0$\\

\hline
   $\Phi^{\alpha\beta}_{1} $ & $b_1+ b_2$      & $(1,1)$ & $0$ & $0$ &
   $0$ & $0$ & $0$ & $-{1\over2}$ & ${1\over2}$ & $0$ &
   $(1,2,1,2,1)$ & $~~0$  \\
   $\bar\Phi^{\alpha\beta}_{1} $ & $\alpha + \beta$  &  $(1,1)$ & $0$ & $0$ &
  $0$ & $0$ & $0$ & ${1\over2}$ & $-{1\over2}$ & $0$ &
   $(1,2,1,2,1)$ & $~~0$  \\
   $\Phi^{\alpha\beta}_{2} $ & $ $  & $(1,1)$ & $0$ & $0$ &
   $0$ & $0$ & $0$ & $-{1\over2}$ & ${1\over2}$ & $0$ &
   $(2,1,2,1,1)$ & $~~0$  \\
   $\bar \Phi^{\alpha\beta}_{2}$ & $ $           & $(1,1)$ & $0$ & $0$ &
   $0$ & $0$ & $0$ & ${1\over2}$ & $-{1\over2}$ & $0$ &
   $(2,1,2,1,1)$ & $~~0$  \\
  
\hline
   $V_1 $ & $b_1+ 2 \gamma$      & $(1,1)$ & $0$ & $0$ &
   $0$ & $-{1\over2}$ & $1\over2$ & ${1\over2}$ & $0$ & $0$ &
   $(1,1,1,1,6)$ & $~~0$  \\
   $V_2 $ &           &  $(1,1)$ & $0$ & $0$ &
   $0$ & $-{1\over2}$ & $1\over2$ & $-{1\over2}$ & $0$ & $0$ &
   $(1,1,1,1,1)$ & $-2$  \\
   $V_3 $ & $ $           & $(1,1)$ & $0$ & $0$ &
   $0$ & $-{1\over2}$ & $1\over2$ & $-{1\over2}$ & $0$ & $0$ &
   $(1,1,1,1,1)$ & $~~2$  \\
   \hline
   $V_4 $ & $b_2+ 2 \gamma$      & $(1,1)$ & $0$ & $0$ &
   $-{1\over2}$ & $0$ & $1\over2$ & $0$ & $-{1\over2}$ & $0$ &
   $(1,1,1,1,6)$ & $~~0$  \\
   $V_5 $ &            & $(1,1)$ & $0$ & $0$ &
   $-{1\over2}$ & $0$ & $1\over2$ & $0$ & ${1\over2}$ & $0$ &
   $(1,1,1,1,1)$ & $-2$  \\
   $V_6 $ &            & $(1,1)$ & $0$ & $0$ &
   $-{1\over2}$ & $0$ & $1\over2$ & $0$ & ${1\over2}$ & $0$ &
   $(1,1,1,1,1)$ & $~~2$  \\
\hline
   $V_7 $ & $b_3+ 2 \gamma$      & $(1,1)$ & $0$ & $0$ &
   $-{1\over2}$ & $-{1\over2}$ & $0$ & $0$ & $0$ & $1\over2$ &
   $(1,1,1,1,6)$ & $~~0$  \\
   $V_8 $ &            & $(1,1)$ & $0$ & $0$ &
   $-{1\over2}$ & $-{1\over2}$ & $0$ & $0$ & $0$ & $-{1\over2}$ &
   $(1,1,1,1,1)$ & $-2$  \\
   $V_9 $ &            & $(1,1)$ & $0$ & $0$ &
   $-{1\over2}$ & $-{1\over2}$ & $0$ & $0$ & $0$ & $-{1\over2}$ &
   $(1,1,1,1,1)$ & $~~2$  \\
\hline
  
   $V_{10} $ & $1+b_2+ $      & $(1,1)$ & $0$ & $0$ &
   $0$ & $-{1\over2}$ & $1\over2$ & $-{1\over2}$ & $0$ & $0$ &
   $(1,2,2,1,1)$ & $~~0$   \\
 $V_{11} $ & $b_3+2\gamma $      & $(1,1)$ & $0$ & $0$ &
   $0$ & $-{1\over2}$ & $1\over2$ & $-{1\over2}$ & $0$ & $0$ &
   $(2,1,1,2,1)$ & $~~0$   \\
\hline
   $V_{12} $ & $1+b_1+ $      & $(1,1)$ & $0$ & $0$ &
   $-{1\over2}$ & $0$ & $1\over2$ & $0$ & $1\over2$ & $0$ &
   $(2,1,1,2,1)$ & $~~0$   \\
 $V_{13} $ & $b_3+2\gamma $      & $(1,1)$ & $0$ & $0$ &
   $-{1\over2}$ & $0$ & $1\over2$ & $0$ & ${1\over2}$ & $0$ &
   $(1,2,2,1,1)$ & $~~0$   \\
\hline
   $V_{14}$ & $1+b_1+ $      & $(1,1)$ & $0$ & $0$ &
   $-{1\over2}$ & $-{1\over2}$ & $0$ & $0$ & $0$ & $-{1\over2}$ &
   $(2,1,1,2,1)$ & $~~0$   \\
 $V_{15} $ & $b_2+2\gamma $      & $(1,1)$ & $0$ & $0$ &
   $-{1\over2}$ & $-{1\over2}$ & $0$ & $0$ & $0$ & $-{1\over2}$ &
   $(1,2,2,1,1)$ & $~~0$   \\
\hline
 $H_1 $ & $b_1+ \alpha$    & $(1,2)$ & $0$ & $0$ &
   $0$ & $0$ & $0$ &$0$ & $-{1\over2}$  & ${1\over2}$ &
  $(2,1,1,1,1)$ & $~~0$   \\
 $\bar{H}_1 $ & $ $      & $(1,2)$ & $0$ & $0$ &
   $0$ & $0$ & $0$ & $0$ &${1\over2}$  & $-{1\over2}$ &
   $(2,1,1,1,1)$& $~~0$   \\
 \hline
   $H_2 $ &  $b_2+ \beta$       & $(1,2)$ & $0$ & $0$ &
   $0$ & $0$ & $0$ & ${1\over2}$ &$0$ & $-{1\over2}$ &
   $(1,1,2,1,1)$  & $~~0$   \\
 $\bar{H}_2 $ & $ $      & $(1,2)$ & $0$ & $0$ &
   $0$ & $0$ & $0$ & $-{1\over2}$  &$0$ & ${1\over2}$ &
$(1,1,2,1,1)$  & $~~0$   \\

\hline
\end{tabular}
\label{matter2}
\end{eqnarray*}
Table 1. continued.
\end{table}


\newpage
\renewcommand{\baselinestretch}{1.3}
\begin{table}
\begin{eqnarray*}
\begin{tabular}{|c|c|c|rrrrrrrr|c|c|}
\hline
  $F$ & SEC & $SU(3)\times$&$Q_{C}$ & $Q_L$ & $Q_1$ &
   $Q_2$ & $Q_3$ & $Q_{4}$ & $Q_{5}$ & $Q_6$ &
   $SU(2)_{1,..,4}$ & $Q_{H_1}$  \\
   $$ & $$ & $SU(2)$ & $$ & $$ & $$ & $$ & $$ & $$ & $$ & $$ & $\times 
							SU(4)_{H_1}$  & $$  \\
\hline
   $H_3 $ & $b_3\pm \gamma$      & $(\bar{3},1)$ & $1\over4$ & $-{1\over2}$ &
   $1\over4$ & $1\over4$ & $-{1\over4}$ & $0$ & $0$ & $-{1\over2}$ &
   $(1,1,1,1,1)$ & $~~1$  \\
   $H_{4} $ &            & $(1,2)$ & $-{3\over4}$ & $1\over2$ &
   $1\over4$ & $1\over4$ & $-{1\over4}$ & $0$ & $0$ & $-{1\over2}$ &
   $(1,1,1,1,1)$ & $~~1$  \\
   $H_{5} $ &            & $(1,1)$ & $-{3\over4}$ & $-{1\over2}$ &
   $-{3\over4}$ & $1\over4$ & $-{1\over4}$ & $0$ & $0$ & $-{1\over2}$ &
   $(1,1,1,1,1)$ & $~~1$  \\
   $H_{6}$  &            & $(1,1)$ & $-{3\over4}$ & $-{1\over2}$ &
   $1\over4$ & $-{3\over4}$ & $-{1\over4}$ & $0$ & $0$ & $-{1\over2}$ &
   $(1,1,1,1,1)$ & $~~1$  \\ 
$H_{7} $ &            & $(1,1)$ & $-{3\over4}$ & $-{1\over2}$ &
   $1\over4$ & $1\over4$ & $3\over4$ & $0$ & $0$ & $-{1\over2}$ &
   $(1,1,1,1,1)$ & $~~1$  \\
   $\bar{H}_{3} $ & $ $      & $( 3,1)$ & $-{1\over4}$ & $1\over2$ &
   $-{1\over4}$ & $-{1\over4}$ & $1\over4$ & $0$ & $0$ & ${1\over2}$ &
   $(1,1,1,1,1)$ & $-1$  \\
  $\bar{H}_{4} $ &            & $(1,2)$ & $3\over4$ & $-{1\over2}$ &
   $-{1\over4}$ & $-{1\over4}$ & $1\over4$ & $0$ & $0$ & ${1\over2}$ &
   $(1,1,1,1,1)$ & $-1$  \\
   $\bar{H}_{5} $ &            & $(1,1)$ & $3\over4$ & $1\over2$ &
   ${3\over4}$ & $-{1\over4}$ & $1\over4$ & $0$ & $0$ & ${1\over2}$ &
   $(1,1,1,1,1)$ & $-1$  \\
   $\bar{H}_{6} $ &            & $(1,1)$ & $3\over4$ & $1\over2$ &
   $-{1\over4}$ & ${3\over4}$ & $1\over4$ & $0$ & $0$ & ${1\over2}$ &
   $(1,1,1,1,1)$ & $-1$  \\ 
$\bar{H}_{7} $ &            & $(1,1)$ & $3\over4$ & $1\over2$ &
   $-{1\over4}$ & $-{1\over4}$ & $-{3\over4}$ & $0$ & $0$ & ${1\over2}$ &
   $(1,1,1,1,1)$ & $-1$  \\
\hline
$H_{8} $ & $b_2+ b_3 $      & $(1,1)$ & $3\over4$ & $-{1\over2}$ &
   $-{1\over4}$ & $-{1\over4}$ & $1\over4$ & $1\over2$ & $0$ & $0$ &
   $(1,1,2,1,1)$ & $-1$  \\
   $\bar{H}_{8} $ &  $\beta \pm \gamma$ & $(1,1)$ & $-{3\over4}$ & $1\over2$ &
   $1\over4$ & $1\over4$ & $-{1\over4}$ & $-{1\over2}$ & $0$ & $0$ &
   $(1,1,2,1,1)$ & $~~1$  \\
\hline
 $H_{9} $ & $1+ b_1 $      & $(1,1)$ & $3\over4$ & $-{1\over2}$ &
   $-{1\over4}$ & $-{1\over4}$ & $1\over4$ & ${1\over2}$ & $0$ & $0$ &
   $(1,2,1,1,1)$ & $~~1$  \\
   $\bar{H}_{9} $ &     $ +  \beta \pm \gamma$  & $(1,1)$ & $-{3\over4}$ & 
								${1\over2}$ &
   $1\over4$ & $1\over4$ & $-{1\over4}$ & $-{1\over2}$ & $0$ & $0$ &
   $(1,2,1,1,1)$ & $-1$  \\
\hline
$H_{10} $ & $b_1+ b_3 $      & $(1,1)$ & $-{3\over4}$ & $1\over2$ &
   $1\over4$ & $1\over4$ & $-{1\over4}$ & $0$ & $-{1\over2}$ & $0$ &
   $(2,1,1,1,1)$ & $~~1$  \\
   $\bar{H}_{10} $ &       $ \alpha \pm \gamma $   & $(1,1)$ & $3\over4$ & 
								$-{1\over2}$ &
   $-{1\over4}$ & $-{1\over4}$ & $1\over4$ & $0$ & $1\over2$ & $0$ &
   $(2,1,1,1,1)$ & $-1$  \\
\hline
 $H_{11} $ & $1+ b_2+ $      & $(1,1)$ & $-{3\over4}$ & $1\over2$ &
   $1\over4$ & $1\over4$ & $-{1\over4}$ & $0$ & $-{1\over2}$ & $0$ &
   $(1,1,1,2,1)$ & $-1$  \\
   $\bar{H}_{11} $ &     $ +  \alpha \pm \gamma$    & $(1,1)$ & $3\over4$ & 
								$-{1\over2}$ &
   $-{1\over4}$ & $-{1\over4}$ & $1\over4$ & $0$ & $1\over2$ & $0$ &
   $(1,1,1,2,1)$ & $~~1$  \\
\hline
   $H_{12} $ & $1+ b_3+\alpha $      & $(1,1)$ & $3\over4$ & $1\over2$ &
   $1\over4$ & $1\over4$ & $1\over4$ & $-{1\over2}$ & ${1\over2}$ & $0$ &
   $(1,1,1,1,4)$ & $~~0$  \\
   $H_{13} $ &      $+  \beta \pm \gamma $   & $(1,1)$ & $-{3\over4}$ & 
								$-{1\over2}$ &
   $-{1\over4}$ & $-{1\over4}$ & $-{1\over4}$ & $-{1\over2}$ & $1\over2$ &$0$ &
   $(1,1,1,1,\bar 4)$ & $~~0$  \\
\hline
   $H_{14} $ & $1+ b_2+\alpha $      & $(1,1)$ & $3\over4$ & $1\over2$ &
   $1\over4$ & $-{1\over4}$ & $-{1\over4}$ & $-{1\over2}$ & $0$ &$-{1\over2}$ &
   $(1,1,1,1,4)$ & $~~0$  \\
   $H_{15} $ &     $ +  \beta \pm \gamma  $  & $(1,1)$ & $-{3\over4}$ & 
								$-{1\over2}$ &
   $-{1\over4}$ & $1\over4$ & $1\over4$ & $-{1\over2}$ & $0$ & $-{1\over2}$ &
   $(1,1,1,1,\bar 4)$ & $~~0$  \\
\hline
  $H_{16} $ & $1+ b_1+\alpha $      & $(1,1)$ & $3\over4$ & $1\over2$ &
   $-{1\over4}$ & $1\over4$ & $-{1\over4}$ & $0$ & ${1\over2}$ & $-{1\over2}$ &
   $(1,1,1,1,4)$ & $~~0$  \\
   $H_{17} $ &     $ +  \beta \pm \gamma  $  & $(1,1)$ & $-{3\over4}$ & 
								$-{1\over2}$ &
   ${1\over4}$ & $-{1\over4}$ & ${1\over4}$ & $0$ & ${1\over2}$ &$-{1\over2}$ &
   $(1,1,1,1,\bar 4)$ & $~~0$  \\
\hline
\end{tabular}
\label{matter3}
\end{eqnarray*}
Table 1. continued.
\end{table}

\begin{flushleft}
\begin{table}
\begin{tabular}{|r||r||rrrrrrrrrrrrrr|}
\hline 
\hline
FD &$\frac{Q^{(A)}}{15}$& $\Phi_{46}$ & $\Phi^{'}_{45}$ & 
	$\bar{\Phi}^{'}_{56}$&$V_3$&$V_2$&
	$V_6$&$V_5$&$V_9$&$V_8$&$N^{c}_{1}$&$N^{c}_{2}$&
			$N^{c}_{3}$&$\Phi^{'}_{46}$    &$\Phi_{45}$\\
       &  &$\bar{\Phi}_{56}$&$e^c_1$& $e^c_2$& $e^c_3$ &$H_7$& $H_6$ & 
						$H_5$ &&&&&&&\\
\hline
\hline
${\cal{D}}^{'}_1$&   1&  1& -2&  1&  0&  0&  0&  0&  0&  3&  0&  0&  0&  
				0&  0\\&&   0& 0&  0&  3&  2&  2&  2&&&&&&&\\
${\cal{D}}^{'}_2$&   2&  2& -1& -1&  0&  6&  0&  0&  0&  0&  0&  0&  0&
				0&  0\\&&   0& 0&  0&  6&  1&  4&  7&&&&&&&\\
${\cal{D}}^{'}_3$&   2& -1& -1&  2&  0&  0&  0&  6&  0&  0&  0&  0&  0&
			  	0&  0\\&&   0& 0&  0&  6&  1&  7&  4&&&&&&&\\
${\cal{D}}^{'}_4$&   0&  1& -1&  0&  0&  0&  0&  0&  0&  0&  0&  0&  0&
				0&  0\\&&   0& 0&  2& -2&  1& -1&  0&&&&&&&\\
${\cal{D}}^{'}_5$&   0& -1&  1&  0&  0&  0&  0&  0&  0&  0&  0&  2& -2&  
				0&  0\\&&   0& 0&  0&  0&  1& -1&  0&&&&&&&\\
${\cal{D}}^{'}_6$&   0&  0& -1&  1&  0&  0&  0&  0&  0&  0&  0&  0&  0&  
				0&  0\\&&   0& 2&  0& -2&  1&  0& -1&&&&&&&\\
${\cal{D}}^{'}_7$&   0&  1&  0& -1&  0&  0&  0&  0&  0&  0&  0&  0&  0&  
				0&  1\\&&   0& 0&  0&  0&  0&  0&  0&&&&&&&\\
${\cal{D}}^{'}_8$&   0&  1& -1&  0&  0&  0&  0&  0&  0&  0&  0&  0&  0&  
				0&  0\\&&   1& 0&  0&  0&  0&  0&  0&&&&&&&\\
${\cal{D}}^{'}_9$&   0&  0& -1&  1&  0&  0&  0&  0&  0&  0&  0&  0&  0&  
				1&  0\\&&   0& 0&  0&  0&  0&  0&  0&&&&&&&\\
${\cal{D}}^{'}_{10}$&0&  0&  0&  0&  0&  0&  0&  0&  1&  0&  0&  0&  0&  
				0&  0\\&&   0& 0&  0& -1&  0& -1& -1&&&&&&&\\
${\cal{D}}^{'}_{11}$&0&  0&  1& -1&  0&  0&  0&  0&  0&  0&  2&  0& -2&  
				0&  0\\&&   0& 0&  0&  0&  1&  0& -1&&&&&&&\\
${\cal{D}}^{'}_{12}$&0& -1&  1&  0&  0&  0&  2&  0&  0&  0&  0&  0&  0&  
				0&  0\\&&   0& 0&  0& -2& -1& -1&  -2&&&&&&&\\
${\cal{D}}^{'}_{13}$&0&  0&  1& -1&  2&  0&  0&  0&  0&  0&  0&  0&  0&  
				0&  0\\&&   0& 0&  0& -2& -1& -2& -1&&&&&&&\\
\hline
\hline
\end{tabular}
Table 2.a. $D$-Flat direction basis of non-Abelian singlet fields.

Column 2 specifies the anomalous charge and columns 3 through
16 specify the norm-square VEV components of each basis direction.
The six fields $e^c_i$ and $H_{5,6,7}$ carry hypercharge, 
the remaining do not.
A negative component indicates the vector partner of a field (if it exists) 
must take on VEV rather than the field.
\end{table}
\end{flushleft}
\hfill\vfill\newpage
 
\begin{flushleft}
\begin{table}
\begin{tabular}{|r||r|}
\hline 
\hline
FD & VEV\\
\hline
\hline
${\cal{D}}^{'}_1$& $V_8$\\
${\cal{D}}^{'}_2$& $V_2$\\
${\cal{D}}^{'}_3$& $V_5$\\
${\cal{D}}^{'}_4$& $e^c_2$\\
${\cal{D}}^{'}_5$& $N^{c}_{2}$\\
${\cal{D}}^{'}_6$& $e^c_1$\\
${\cal{D}}^{'}_7$& $\Phi_{45}$\\
${\cal{D}}^{'}_8$& $\bar{\Phi}_{56}$\\
${\cal{D}}^{'}_9$& $\Phi^{'}_{46}$\\
${\cal{D}}^{'}_{10}$& $V_9$\\
${\cal{D}}^{'}_{11}$& $N^{c}_{1}$\\
${\cal{D}}^{'}_{12}$& $V_6$\\
${\cal{D}}^{'}_{13}$& $V_3$\\
\hline
\hline
\end{tabular}
\\
Table 2.b. Unique VEV associated with each non-Abelian singlet field 
$D$-Flat basis direction.
\end{table}
\end{flushleft}

\begin{flushleft}
\begin{table}
\begin{tabular}{|r||r||rrrrrrrrrrrrrr|}
\hline 
\hline
FD    &$\frac{Q^{(A)}}{15}$& $\Phi_{46}$ & $\Phi^{'}_{45}$ & $\bar{\Phi}^{'}_{56}$	&
	$V_3$&$V_2$&
	$V_6$&$V_5$&$V_9$&$V_8$&$N^{c}_{1}$&$N^{c}_{2}$&$N^{c}_{3}$&
  	$\Phi^{'}_{46}$      &$\Phi_{45}$\\
& &$\bar{\Phi}_{56}$&$e^c_1$& $e^c_2$& $e^c_3$ &$H_7$& $H_6$ & $H_5$ &&&&&&&\\
& &$\bar{\Phi}^{\alpha\beta}_{1,2}$&$H_{11}$&
		$H_{10}$ & $\bar{H}_{9}$& $\bar{H}_{8}$ & $H_{16}$
& $H_{14}$ & $H_{12}$ & $V_{1}$ & $V_{4}$ & $V_{7}$ & $H_{17}$ 
		                       & $H_{15}$ & $H_{13}$ \\
& & $V_{15}$& $V_{14}$ & $V_{13}$ & $V_{12}$ & $V_{10}$ & $V_{11}$&&&&&&&&\\
&& $Q_1$ & $Q_2$ & $Q_3$ & $d^c_1$ & $d^c_2$ & $d^c_3$ & $u^c_1$ & $u^c_2$ & 
$u^c_3$ & $H_3$ & $h$   & $L_1$ & $L_2$   & $L_3 $  \\
&& $H_4$&$\bar{H}_1$ & $\bar{H}_2$ &&&&&&&&&&&\\                 
\hline
\hline
${\cal{D}}_{1}$&   -2& 4& 1& 1& 0& 0& 0& 0& 0& 0& 6& 0& 0& 0& 0\\&& 
                       0& 0& 0&-6&-1&-4&-7&  &  &  &  &  &  &  \\&&
                       0& 0& 0& 0& 0& 0& 0& 0& 0& 0& 0& 0& 0& 0\\&& 
                       0& 0& 0& 0& 0& 0&  &  &  &  &  &  &  &  \\&& 
                       0& 0& 0& 0& 0& 0& 0& 0& 0& 0& 0& 0& 0& 0\\&&
                      12& 0& 0&  &  &  &  &  &  &  &  &  &  &  \\ 
                        
${\cal{D}}_{2}$&   -2& 1& 1& 4& 0& 0& 0& 0& 0& 0& 0& 6& 0& 0& 0\\&&
                       0& 0& 0&-6&-1&-7&-4&  &  &  &  &  &  &  \\&&  
                       0& 0& 0& 0& 0& 0& 0& 0& 0& 0& 0& 0& 0& 0\\&& 
                       0& 0& 0& 0& 0& 0&  &  &  &  &  &  &  &  \\&& 
                       0& 0& 0& 0& 0& 0& 0& 0& 0& 0& 0& 0& 0& 0\\&&
                      12& 0& 0&  &  &  &  &  &  &  &  &  &  &  \\ 
                       
${\cal{D}}_{3}$&   -1& 2&-1& 2& 0& 0& 0& 0& 0& 0& 0& 0& 3& 0& 0\\&& 
                       0& 0& 0&-3&-2&-2&-2&  &  &  &  &  &  &  \\&& 
                       0& 0& 0& 0& 0& 0& 0& 0& 0& 0& 0& 0& 0& 0\\&& 
                       0& 0& 0& 0& 0& 0&  &  &  &  &  &  &  &  \\&& 
                       0& 0& 0& 0& 0& 0& 0& 0& 0& 0& 0& 0& 0& 0\\&&
                       6& 0& 0&  &  &  &  &  &  &  &  &  &  &  \\
                       
${\cal{D}}_{4}$&   -1& 2&-1& 8& 0& 0& 0& 0& 0& 0& 0& 0& 0& 0& 0\\&&
                       0& 0& 0&-6& 4& 1&-5&  &  &  &  &  &  &  \\&&  
                       0& 0& 0& 0& 0&12& 0& 0& 0& 0& 0& 0& 0& 0\\&& 
                       0& 0& 0& 0& 0& 0&  &  &  &  &  &  &  &  \\&& 
                       0& 0& 0& 0& 0& 0& 0& 0& 0& 0& 0& 0& 0& 0\\&&
                       0& 0& 0&  &  &  &  &  &  &  &  &  &  &  \\
                       
${\cal{D}}_{5}$&   -1& 8&-1& 2& 0& 0& 0& 0& 0& 0& 0& 0& 0& 0& 0\\&&
                       0& 0& 0&-6& 4&-5& 1&  &  &  &  &  &  &  \\&&  
                       0& 0& 0& 0& 0& 0&12& 0& 0& 0& 0& 0& 0& 0\\&& 
                       0& 0& 0& 0& 0& 0&  &  &  &  &  &  &  &  \\&& 
                       0& 0& 0& 0& 0& 0& 0& 0& 0& 0& 0& 0& 0& 0\\&&
                       0& 0& 0&  &  &  &  &  &  &  &  &  &  &  \\
                       
${\cal{D}}_{6}$&   -1& 2& 5& 2& 0& 0& 0& 0& 0& 0& 0& 0& 0& 0& 0\\&&
                       0& 0& 0&-6&-2& 1& 1&  &  &  &  &  &  &  \\&&  
                       0& 0& 0& 0& 0& 0& 0&12& 0& 0& 0& 0& 0& 0\\&& 
                       0& 0& 0& 0& 0& 0&  &  &  &  &  &  &  &  \\&& 
                       0& 0& 0& 0& 0& 0& 0& 0& 0& 0& 0& 0& 0& 0\\&&
                       0& 0& 0&  &  &  &  &  &  &  &  &  &  &  \\
                       
\hline
\hline
\end{tabular}
Table 3.a $D$-Flat direction basis of all fields.
\end{table}
\end{flushleft}

\begin{flushleft}
\begin{table}
\begin{tabular}{|r||r||rrrrrrrrrrrrrr|}
\hline 
\hline
FD  &$\frac{Q^{(A)}}{15}$& $\Phi_{46}$ & $\Phi^{'}_{45}$ & $\bar{\Phi}^{'}_{56}$&$V_3$&$V_2$&
$V_6$&$V_5$&$V_9$&$V_8$&$N^{c}_{1}$&$N^{c}_{2}$&$N^{c}_{3}$&$\Phi^{'}_{46}$  
				&$\Phi_{45}$\\
&         &$\bar{\Phi}_{56}$&$e^c_1$& $e^c_2$& $e^c_3$ &$H_7$& $H_6$ & 
						$H_5$ &&&&&&&\\
& & $\bar{\Phi}^{\alpha\beta}_{1,2}$ & $H_{11}$ & $H_{10}$ & $\bar{H}_{9}$ & 
	$\bar{H}_{8}$ & $H_{16}$ & $H_{14}$ & $H_{12}$ & $V_{1}$ & $V_{4}$ & 
	$V_{7}$ & $H_{17}$ & $H_{15}$ & $H_{13}$ \\
& & $V_{15}$& $V_{14}$ & $V_{13}$ & $V_{12}$ & $V_{10}$ & $V_{11}$&&&&&&&&\\
&& $Q_1$ & $Q_2$ & $Q_3$ & $d^c_1$ & $d^c_2$ & $d^c_3$ & $u^c_1$ & $u^c_2$ & 
$u^c_3$  & $H_3$ & $h$   & $L_1$ & $L_2$   & $L_3 $  \\
&& $H_4$&$\bar{H}_1$ & $\bar{H}_2$ &&&&&&&&&&&\\                  
\hline
\hline
${\cal{D}}_{7}$&   -1& 1& 0& 1& 0& 0& 0& 0& 0& 0& 0& 0& 0& 0& 0\\&&
                       0& 0& 0&-2&-1&-2&-2&  &  &  &  &  &  &  \\&&  
                       0& 0& 0& 0& 0& 0& 0& 0& 0& 0& 0& 0& 0& 0\\&& 
                       0& 0& 0& 0& 0& 0&  &  &  &  &  &  &  &  \\&& 
                       0& 0& 0& 0& 0& 0& 0& 0& 0& 3& 0& 0& 0& 0\\&&
                       2& 0& 0&  &  &  &  &  &  &  &  &  &  &  \\
                       
${\cal{D}}_{8}$&   -1&-1&-1& 2& 0& 0& 0& 0& 0& 0& 0& 0& 0& 0& 0\\&&
                       0& 0& 0&-6&-2&-5&-5&  &  &  &  &  &  &  \\&&  
                       0& 0& 0&-6& 0& 0& 0& 0& 0& 0& 0& 0& 0& 0\\&& 
                       0& 0& 0& 0& 0& 0&  &  &  &  &  &  &  &  \\&& 
                       0& 0& 0& 0& 0& 0& 0& 0& 0& 0& 0& 0& 0& 0\\&&
                       6& 0& 0&  &  &  &  &  &  &  &  &  &  &  \\
                       
${\cal{D}}_{9}$&   -1&-1& 2& 2& 0& 0& 0& 0& 0& 0& 0& 0& 0& 0& 0\\&&
                       0& 0& 0&-6&-2&-5&-5&  &  &  &  &  &  &  \\&&  
                       0&-6& 0& 0& 0& 0& 0& 0& 0& 0& 0& 0& 0& 0\\&& 
                       0& 0& 0& 0& 0& 0&  &  &  &  &  &  &  &  \\&& 
                       0& 0& 0& 0& 0& 0& 0& 0& 0& 0& 0& 0& 0& 0\\&&
                       6& 0& 0&  &  &  &  &  &  &  &  &  &  &  \\
                      
${\cal{D}}_{10}$&   0& 1& 0&-1& 0& 0& 0& 0& 0& 0& 0& 0& 0& 0& 1\\&&
                       0& 0& 0& 0& 0& 0& 0&  &  &  &  &  &  &  \\&&  
                       0& 0& 0& 0& 0& 0& 0& 0& 0& 0& 0& 0& 0& 0\\&& 
                       0& 0& 0& 0& 0& 0&  &  &  &  &  &  &  &  \\&& 
                       0& 0& 0& 0& 0& 0& 0& 0& 0& 0& 0& 0& 0& 0\\&&
                       0& 0& 0&  &  &  &  &  &  &  &  &  &  &  \\
                       
${\cal{D}}_{11}$&   0& 0& 1&-1& 0& 0& 0& 0& 0& 0& 0& 0& 0& 0& 0\\&&
                       0& 0& 0& 0& 0& 0& 0&  &  &  &  &  &  &  \\&&  
                       0& 0& 0& 0& 0& 0& 0& 0& 0& 0& 0& 0& 0& 0\\&& 
                       0& 0& 0& 0& 0& 0&  &  &  &  &  &  &  &  \\&& 
                       0& 0& 0& 0& 0& 0& 0& 0& 0& 0& 0& 0& 0& 0\\&&
                       0& 0& 2&  &  &  &  &  &  &  &  &  &  &  \\
                       
${\cal{D}}_{12}$&   0& 0& 0& 1& 0& 0& 0& 0& 0& 0& 0& 0& 0& 0& 0\\&&
                       0& 0& 0& 0& 0& 0& 0&  &  &  &  &  &  &  \\&&
                       0& 0& 0& 0& 0& 0& 0& 0& 0& 0& 0& 0& 0& 0\\&& 
                       0& 0& 0& 0& 0& 0&  &  &  &  &  &  &  &  \\&& 
                       0& 0& 0& 0& 0& 0& 0& 0& 0& 0& 0& 0& 0& 0\\&& 
                       0& 2& 0&  &  &  &  &  &  &  &  &  &  &  \\
                         
\hline
\hline
\end{tabular}
Table 3.a continued.
\end{table}
\end{flushleft}

\begin{flushleft}
\begin{table}
\begin{tabular}{|r||r||rrrrrrrrrrrrrr|}
\hline 
\hline
FD    &$\frac{Q^{(A)}}{15}$& $\Phi_{46}$ & $\Phi^{'}_{45}$ & $\bar{\Phi}^{'}_{56}$&$V_3$&
$V_2$& $V_6$&$V_5$&$V_9$&$V_8$&$N^{c}_{1}$&$N^{c}_{2}$&$N^{c}_{3}$&$
\Phi^{'}_{46}$      &$\Phi_{45}$\\
& &$\bar{\Phi}_{56}$&$e^c_1$& $e^c_2$& $e^c_3$ &$H_7$& $H_6$ & $H_5$ &&&&&&&\\
& & $\bar{\Phi}^{\alpha\beta}_{1,2}$ & $H_{11}$ & $H_{10}$ & $\bar{H}_{9}$ &
$\bar{H}_{8}$ & $H_{16}$ & $H_{14}$ & $H_{12}$ & $V_{1}$ & $V_{4}$ & $V_{7}$ 
				& $H_{17}$ & $H_{15}$ & $H_{13}$ \\
& & $V_{15}$& $V_{14}$ & $V_{13}$ & $V_{12}$ & $V_{10}$ & $V_{11}$&&&&&&&&\\
&& $Q_1$ & $Q_2$ & $Q_3$ & $d^c_1$ & $d^c_2$ & $d^c_3$ & $u^c_1$ & $u^c_2$ &
$u^c_3$ & $H_3$ & $h$   & $L_1$ & $L_2$   & $L_3 $  \\
&& $H_4$&$\bar{H}_1$ & $\bar{H}_2$ &&&&&&&&&&&\\                   
\hline
\hline
${\cal{D}}_{13}$   &0& 0& 0&-1& 0& 0& 0& 0& 0& 0& 0& 0& 0& 0& 0\\&&
                       0& 0& 0& 0& 0& 0& 0&  &  &  &  &  &  &  \\&& 
                       0& 0& 2& 0& 0& 0& 0& 0& 0& 0& 0& 0& 0& 0\\&& 
                       0& 0& 0& 0& 0& 0&  &  &  &  &  &  &  &  \\&& 
                       0& 0& 0& 0& 0& 0& 0& 0& 0& 0& 0& 0& 0& 0\\&&
                      -2& 0& 0&  &  &  &  &  &  &  &  &  &  &  \\
                        
${\cal{D}}_{14}$&   0& 0&-1& 1& 0& 0& 0& 0& 0& 0& 0& 0& 0& 1& 0\\&&
                       0& 0& 0& 0& 0& 0& 0&  &  &  &  &  &  &  \\&&
                       0& 0& 0& 0& 0& 0& 0& 0& 0& 0& 0& 0& 0& 0\\&& 
                       0& 0& 0& 0& 0& 0&  &  &  &  &  &  &  &  \\&& 
                       0& 0& 0& 0& 0& 0& 0& 0& 0& 0& 0& 0& 0& 0\\&&
                       0& 0& 0&  &  &  &  &  &  &  &  &  &  &  \\
                         
${\cal{D}}_{15}$&   0& 0&-1& 0& 0& 0& 0& 0& 0& 0& 0& 0& 0& 0& 0\\&&
                       0& 0& 0& 0& 0& 0& 0&  &  &  &  &  &  &  \\&&  
                       2& 0& 0& 0& 0& 0& 0& 0& 0& 0& 0& 0& 0& 0\\&& 
                       0& 0& 0& 0& 0& 0&  &  &  &  &  &  &  &  \\&& 
                       0& 0& 0& 0& 0& 0& 0& 0& 0& 0& 0& 0& 0& 0\\&&
                       0& 0& 0&  &  &  &  &  &  &  &  &  &  &  \\
                       
${\cal{D}}_{16}$&   0& 1&-1& 0& 0& 0& 0& 0& 0& 0& 0& 0& 0& 0& 0\\&&
                       1& 0& 0& 0& 0& 0& 0&  &  &  &  &  &  &  \\&& 
                       0& 0& 0& 0& 0& 0& 0& 0& 0& 0& 0& 0& 0& 0\\&& 
                       0& 0& 0& 0& 0& 0&  &  &  &  &  &  &  &  \\&& 
                       0& 0& 0& 0& 0& 0& 0& 0& 0& 0& 0& 0& 0& 0\\&&
                       0& 0& 0&  &  &  &  &  &  &  &  &  &  &  \\
                        
${\cal{D}}_{17}$&   0& 0& 1&-1& 0& 0& 0& 0& 0& 0& 0& 0& 0& 0& 0\\&&
                       0& 0& 0& 0& 0& 0& 0&  &  &  &  &  &  &  \\&& 
                       0& 0& 0& 0& 2& 0& 0& 0& 0& 0& 0& 0& 0& 0\\&& 
                       0& 0& 0& 0& 0& 0&  &  &  &  &  &  &  &  \\&& 
                       0& 0& 0& 0& 0& 0& 0& 0& 0& 0& 0& 0& 0& 0\\&&
                      -2& 0& 0&  &  &  &  &  &  &  &  &  &  &  \\
                        
\hline
\hline
\end{tabular}
Table 3.a continued.
\end{table}
\end{flushleft}

\begin{flushleft}
\begin{table}
\begin{tabular}{|r||r||rrrrrrrrrrrrrr|}
\hline 
\hline
FD    &$\frac{Q^{(A)}}{15}$& $\Phi_{46}$ & $\Phi^{'}_{45}$ & $\bar{\Phi}^{'}_{56}$&
	$V_3$&$V_2$& $V_6$&$V_5$&$V_9$&$V_8$&$N^{c}_{1}$&$N^{c}_{2}$&$N^{c}_{3}$&
	$\Phi^{'}_{46}$      &$\Phi_{45}$\\
& &$\bar{\Phi}_{56}$&$e^c_1$& $e^c_2$& $e^c_3$ &$H_7$& $H_6$ & $H_5$ &&&&&&&\\
& & $\bar{\Phi}^{\alpha\beta}_{1,2}$ & $H_{11}$ & $H_{10}$ & $\bar{H}_{9}$ & 
$\bar{H}_{8}$ & $H_{16}$ & $H_{14}$ & $H_{12}$ & $V_{1}$ & $V_{4}$ & $V_{7}$ & 
$H_{17}$ & $H_{15}$ & $H_{13}$ \\
& & $V_{15}$& $V_{14}$ & $V_{13}$ & $V_{12}$ & $V_{10}$ & $V_{11}$&&&&&&&&\\
&& $Q_1$ & $Q_2$ & $Q_3$ & $d^c_1$ & $d^c_2$ & $d^c_3$ & $u^c_1$ & $u^c_2$ & 
$u^c_3$ & $H_3$ & $h$   & $L_1$ & $L_2$   & $L_3 $  \\
&& $H_4$&$\bar{H}_1$ & $\bar{H}_2$ &&&&&&&&&&&\\                          
\hline
\hline
${\cal{D}}_{18}$&   0&-1& 1& 0& 0& 0& 2& 0& 0& 0& 0& 0& 0& 0& 0\\&&
                       0& 0& 0&-2&-1&-1&-2&  &  &  &  &  &  &  \\&&  
                       0& 0& 0& 0& 0& 0& 0& 0& 0& 0& 0& 0& 0& 0\\&& 
                       0& 0& 0& 0& 0& 0&  &  &  &  &  &  &  &  \\&& 
                       0& 0& 0& 0& 0& 0& 0& 0& 0& 0& 0& 0& 0& 0\\&&
                       0& 0& 0&  &  &  &  &  &  &  &  &  &  &  \\
                       
${\cal{D}}_{19}$&   0& 2& 1&-1& 0& 0& 0& 0& 0& 0& 0& 0& 0& 0& 0\\&&
                       0& 0& 0&-2& 1& 0&-3&  &  &  &  &  &  &  \\&& 
                       0& 0& 0& 0& 0& 0& 0& 0& 0& 0& 0& 0& 0& 0\\&& 
                       0& 0& 0& 0& 0& 0&  &  &  &  &  &  &  &  \\&& 
                       6& 0& 0& 0& 0& 0& 0& 0& 0& 0& 0& 0& 0& 0\\&&
                       2& 0& 0&  &  &  &  &  &  &  &  &  &  &  \\

${\cal{D}}_{20}$&   0& 1&-1& 1& 0& 0& 0& 0& 0& 0& 0& 0& 0& 0& 0\\&&
                       0& 0& 0&-1&-1& 0& 0&  &  &  &  &  &  &  \\&& 
                       0& 0& 0& 0& 0& 0& 0& 0& 0& 0& 0& 0& 0& 0\\&& 
                       0& 0& 0& 0& 0& 0&  &  &  &  &  &  &  &  \\&& 
                       0& 0& 3& 0& 0& 0& 0& 0& 0& 0& 0& 0& 0& 0\\&&
                       1& 0& 0&  &  &  &  &  &  &  &  &  &  &  \\

${\cal{D}}_{21}$&   0&-1& 1&-1& 0& 0& 0& 0& 0& 0& 0& 0& 0& 0& 0\\&&
                       0& 0& 0& 1&-2& 0& 0&  &  &  &  &  &  &  \\&& 
                       0& 0& 0& 0& 0& 0& 0& 0& 0& 0& 0& 0& 0& 0\\&& 
                       0& 0& 0& 0& 0& 0&  &  &  &  &  &  &  &  \\&& 
                       0& 0& 0& 0& 0& 0& 0& 0& 3& 0& 0& 0& 0& 0\\&&
                       2& 0& 0&  &  &  &  &  &  &  &  &  &  &  \\
                        
${\cal{D}}_{22}$&   0& 0& 0& 0& 0& 0& 0& 0& 0& 0& 0& 0& 0& 0& 0\\&&
                       0& 0& 0& 0&-1& 0& 0&  &  &  &  &  &  &  \\&& 
                       0& 0& 0& 0& 0& 0& 0& 0& 0& 0& 0& 0& 0& 0\\&& 
                       0& 0& 0& 0& 0& 0&  &  &  &  &  &  &  &  \\&& 
                       0& 0& 0& 0& 0& 0& 0& 0& 0& 0& 1& 0& 0& 0\\&&
                       1& 0& 0&  &  &  &  &  &  &  &  &  &  &  \\
                        
${\cal{D}}_{23}$&   0& 1&-1& 0& 0& 0& 0& 0& 0& 0& 0& 0& 0& 0& 0\\&&
                       0& 0& 2&-2& 1&-1& 0&  &  &  &  &  &  &  \\&&  
                       0& 0& 0& 0& 0& 0& 0& 0& 0& 0& 0& 0& 0& 0\\&& 
                       0& 0& 0& 0& 0& 0&  &  &  &  &  &  &  &  \\&& 
                       0& 0& 0& 0& 0& 0& 0& 0& 0& 0& 0& 0& 0& 0\\&&
                       0& 0& 0&  &  &  &  &  &  &  &  &  &  &  \\
                       
\hline
\hline
\end{tabular}
Table 3.a continued.
\end{table}
\end{flushleft}

\begin{flushleft}
\begin{table}
\begin{tabular}{|r||r||rrrrrrrrrrrrrr|}
\hline 
\hline
FD    &$\frac{Q^{(A)}}{15}$& $\Phi_{46}$ & $\Phi^{'}_{45}$ & $\bar{\Phi}^{'}_{56}$&
	$V_3$&$V_2$& $V_6$&$V_5$&$V_9$&$V_8$&$N^{c}_{1}$&$N^{c}_{2}$&$N^{c}_{3}$&
	$\Phi^{'}_{46}$ &$\Phi_{45}$\\
& &$\bar{\Phi}_{56}$&$e^c_1$& $e^c_2$& $e^c_3$ &$H_7$& $H_6$ & $H_5$ &&&&&&&\\
& & $\bar{\Phi}^{\alpha\beta}_{1,2}$ & $H_{11}$ & $H_{10}$ & $\bar{H}_{9}$ & 
$\bar{H}_{8}$ & $H_{16}$ & $H_{14}$ & $H_{12}$ & $V_{1}$ & $V_{4}$ & $V_{7}$ & 
$H_{17}$ & $H_{15}$ & $H_{13}$ \\
& & $V_{15}$& $V_{14}$ & $V_{13}$ & $V_{12}$ & $V_{10}$ & $V_{11}$&&&&&&&&\\
&& $Q_1$ & $Q_2$ & $Q_3$ & $d^c_1$ & $d^c_2$ & $d^c_3$ & $u^c_1$ & $u^c_2$ & 
$u^c_3$ & $H_3$ & $h$   & $L_1$ & $L_2$   & $L_3 $  \\
&& $H_4$&$\bar{H}_1$ & $\bar{H}_2$ &&&&&&&&&&&\\                           
\hline
\hline
${\cal{D}}_{24}$&   0& 0&-1& 1& 0& 0& 0& 0& 0& 0& 0& 0& 0& 0& 0\\&&
                       0& 2& 0&-2& 1& 0&-1&  &  &  &  &  &  &  \\&&  
                       0& 0& 0& 0& 0& 0& 0& 0& 0& 0& 0& 0& 0& 0\\&& 
                       0& 0& 0& 0& 0& 0&  &  &  &  &  &  &  &  \\&& 
                       0& 0& 0& 0& 0& 0& 0& 0& 0& 0& 0& 0& 0& 0\\&&
                       0& 0& 0&  &  &  &  &  &  &  &  &  &  &  \\
                       
${\cal{D}}_{25}$&   0& 0& 0& 0& 0& 0& 0& 0& 1& 0& 0& 0& 0& 0& 0\\&& 
                       0& 0& 0&-1& 0&-1&-1&  &  &  &  &  &  &  \\&& 
                       0& 0& 0& 0& 0& 0& 0& 0& 0& 0& 0& 0& 0& 0\\&& 
                       0& 0& 0& 0& 0& 0&  &  &  &  &  &  &  &  \\&& 
                       0& 0& 0& 0& 0& 0& 0& 0& 0& 0& 0& 0& 0& 0\\&&
                       0& 0& 0&  &  &  &  &  &  &  &  &  &  &  \\
                        
${\cal{D}}_{26}$&   0& 0& 1&-1& 2& 0& 0& 0& 0& 0& 0& 0& 0& 0& 0\\&&
                       0& 0& 0&-2&-1&-2&-1&  &  &  &  &  &  &  \\&& 
                       0& 0& 0& 0& 0& 0& 0& 0& 0& 0& 0& 0& 0& 0\\&& 
                       0& 0& 0& 0& 0& 0&  &  &  &  &  &  &  &  \\&& 
                       0& 0& 0& 0& 0& 0& 0& 0& 0& 0& 0& 0& 0& 0\\&&
                       0& 0& 0&  &  &  &  &  &  &  &  &  &  &  \\
                        
${\cal{D}}_{27}$&   0&-2&-1& 1& 0& 0& 0& 0& 0& 0& 0& 0& 0& 0& 0\\&&
                       0& 0& 0& 2&-1& 0&-3&  &  &  &  &  &  &  \\&& 
                       0& 0& 0& 0& 0& 0& 0& 0& 0& 0& 0& 0& 0& 0\\&& 
                       0& 0& 0& 0& 0& 0&  &  &  &  &  &  &  &  \\&& 
                       0& 0& 0& 0& 0& 0& 6& 0& 0& 0& 0& 0& 0& 0\\&& 
                       4& 0& 0&  &  &  &  &  &  &  &  &  &  &  \\
                        
${\cal{D}}_{28}$&   0&-1& 1& 2& 0& 0& 0& 0& 0& 0& 0& 0& 0& 0& 0\\&&
                       0& 0& 0&-2& 1&-3& 0&  &  &  &  &  &  &  \\&&
                       0& 0& 0& 0& 0& 0& 0& 0& 0& 0& 0& 0& 0& 0\\&& 
                       0& 0& 0& 0& 0& 0&  &  &  &  &  &  &  &  \\&& 
                       0& 6& 0& 0& 0& 0& 0& 0& 0& 0& 0& 0& 0& 0\\&&
                       2& 0& 0&  &  &  &  &  &  &  &  &  &  &  \\
                         
${\cal{D}}_{29}$&   0& 1&-1&-2& 0& 0& 0& 0& 0& 0& 0& 0& 0& 0& 0\\&&
                       0& 0& 0& 2&-1&-3& 0&  &  &  &  &  &  &  \\&&  
                       0& 0& 0& 0& 0& 0& 0& 0& 0& 0& 0& 0& 0& 0\\&& 
                       0& 0& 0& 0& 0& 0&  &  &  &  &  &  &  &  \\&& 
                       0& 0& 0& 0& 0& 0& 0& 6& 0& 0& 0& 0& 0& 0\\&&
                       4& 0& 0&  &  &  &  &  &  &  &  &  &  &  \\
                       
\hline
\hline
\end{tabular}
Table 3.a continued.
\end{table}
\end{flushleft}

\begin{flushleft}
\begin{table}
\begin{tabular}{|r||r||rrrrrrrrrrrrrr|}
\hline 
\hline
FD    &$\frac{Q^{(A)}}{15}$& $\Phi_{46}$ & $\Phi^{'}_{45}$ & $\bar{\Phi}^{'}_{56}$&$V_3$&
	$V_2$&$V_6$&$V_5$&$V_9$&$V_8$&$N^{c}_{1}$&$N^{c}_{2}$&$N^{c}_{3}$&
	$\Phi^{'}_{46}$      &$\Phi_{45}$\\
& &$\bar{\Phi}_{56}$&$e^c_1$& $e^c_2$& $e^c_3$ &$H_7$& $H_6$ & $H_5$ &&&&&&&\\
& & $\bar{\Phi}^{\alpha\beta}_{1,2}$ & $H_{11}$ & $H_{10}$ & $\bar{H}_{9}$ & 
	$\bar{H}_{8}$ & $H_{16}$ & $H_{14}$ & $H_{12}$ & $V_{1}$ & $V_{4}$ & 
	$V_{7}$ & $H_{17}$ & $H_{15}$ & $H_{13}$ \\
& & $V_{15}$& $V_{14}$ & $V_{13}$ & $V_{12}$ & $V_{10}$ & $V_{11}$&&&&&&&&\\
&& $Q_1$ & $Q_2$ & $Q_3$ & $d^c_1$ & $d^c_2$ & $d^c_3$ & $u^c_1$ & $u^c_2$ & 
$u^c_3$ & $H_3$ & $h$   & $L_1$ & $L_2$   & $L_3 $  \\
&& $H_4$&$\bar{H}_1$ & $\bar{H}_2$ &&&&&&&&&&&\\             
\hline
\hline
${\cal{D}}_{30}$&   1& 4& 1&-2& 0& 0& 0& 0& 0& 0& 0& 0& 0& 0& 0\\&&
                       0& 0& 0& 6&-4& 5&-1&  &  &  &  &  &  &  \\&&  
                       0& 0& 0& 0& 0& 0& 0& 0& 0& 0& 0& 0&12& 0\\&& 
                       0& 0& 0& 0& 0& 0&  &  &  &  &  &  &  &  \\&& 
                       0& 0& 0& 0& 0& 0& 0& 0& 0& 0& 0& 0& 0& 0\\&&
                       0& 0& 0&  &  &  &  &  &  &  &  &  &  &  \\
                       
${\cal{D}}_{31}$&   1& 1&-2&-2& 0& 0& 0& 0& 0& 0& 0& 0& 0& 0& 0\\&&
                       0& 0& 0& 0&-1& 2&-1&  &  &  &  &  &  &  \\&&  
                       0& 0& 0& 0& 0& 0& 0& 0& 0& 6& 0& 0& 0& 0\\&& 
                       0& 0& 0& 0& 0& 0&  &  &  &  &  &  &  &  \\&& 
                       0& 0& 0& 0& 0& 0& 0& 0& 0& 0& 0& 0& 0& 0\\&&
                       0& 0& 0&  &  &  &  &  &  &  &  &  &  &  \\
                       
${\cal{D}}_{32}$&   1&-2& 7&-2& 0& 0& 0& 0& 0& 0& 0& 0& 0& 0& 0\\&&
                       0& 0& 0& 6& 2&-1&-1&  &  &  &  &  &  &  \\&&
                       0& 0& 0& 0& 0& 0& 0& 0& 0& 0& 0& 0& 0&12\\&&
                       0& 0& 0& 0& 0& 0&  &  &  &  &  &  &  &  \\&& 
                       0& 0& 0& 0& 0& 0& 0& 0& 0& 0& 0& 0& 0& 0\\&&
                       0& 0& 0&  &  &  &  &  &  &  &  &  &  &  \\
                         
${\cal{D}}_{33}$&   1& 1& 1&-2& 0& 0& 0& 0& 0& 0& 0& 0& 0& 0& 0\\&&
                       0& 0& 0& 0&-1&-1& 2&  &  &  &  &  &  &  \\&&
                       0& 0& 0& 0& 0& 0& 0& 0& 0& 0& 0& 0& 0& 0\\&& 
                       0& 0& 0& 0& 6& 0&  &  &  &  &  &  &  &  \\&& 
                       0& 0& 0& 0& 0& 0& 0& 0& 0& 0& 0& 0& 0& 0\\&&
                       0& 0& 0&  &  &  &  &  &  &  &  &  &  &  \\
                         
${\cal{D}}_{34}$&   1& 1&-2& 1& 0& 0& 0& 0& 0& 0& 0& 0& 0& 0& 0\\&&
                       0& 0& 0& 0& 2&-1&-1&  &  &  &  &  &  &  \\&&  
                       0& 0& 0& 0& 0& 0& 0& 0& 0& 0& 0& 0& 0& 0\\&& 
                       6& 0& 0& 0& 0& 0&  &  &  &  &  &  &  &  \\&& 
                       0& 0& 0& 0& 0& 0& 0& 0& 0& 0& 0& 0& 0& 0\\&&
                       0& 0& 0&  &  &  &  &  &  &  &  &  &  &  \\
                       
${\cal{D}}_{35}$&   1& 1&-2& 1& 0& 0& 0& 0& 0& 0& 0& 0& 0& 0& 0\\&&
                       0& 0& 0& 0& 2&-1&-1&  &  &  &  &  &  &  \\&& 
                       0& 0& 0& 0& 0& 0& 0& 0& 0& 0& 0& 0& 0& 0\\&& 
                       0& 6& 0& 0& 0& 0&  &  &  &  &  &  &  &  \\&& 
                       0& 0& 0& 0& 0& 0& 0& 0& 0& 0& 0& 0& 0& 0\\&&
                       0& 0& 0&  &  &  &  &  &  &  &  &  &  &  \\
                        
\hline
\hline
\end{tabular}
Table 3.a continued.
\end{table}
\end{flushleft}

\begin{flushleft}
\begin{table}
\begin{tabular}{|r||r||rrrrrrrrrrrrrr|}
\hline 
\hline
FD  &$\frac{Q^{(A)}}{15}$& $\Phi_{46}$ & $\Phi^{'}_{45}$ & $\bar{\Phi}^{'}_{56}$&$V_3$&$V_2$&
$V_6$&$V_5$&$V_9$&$V_8$&$N^{c}_{1}$&$N^{c}_{2}$&$N^{c}_{3}$&$\Phi^{'}_{46}$
      &$\Phi_{45}$\\
      & &$\bar{\Phi}_{56}$&$e^c_1$& $e^c_2$& $e^c_3$ &$H_7$& $H_6$ & 
	$H_5$ &&&&&&&\\
& & $\bar{\Phi}^{\alpha\beta}_{1,2}$ & $H_{11}$ & $H_{10}$ & $\bar{H}_{9}$ & 
	$\bar{H}_{8}$ & $H_{16}$ & $H_{14}$ & $H_{12}$ & $V_{1}$ & $V_{4}$ & 
	$V_{7}$ & $H_{17}$ & $H_{15}$ & $H_{13}$ \\
& & $V_{15}$& $V_{14}$ & $V_{13}$ & $V_{12}$ & $V_{10}$ & $V_{11}$&&&&&&&&\\
&& $Q_1$ & $Q_2$ & $Q_3$ & $d^c_1$ & $d^c_2$ & $d^c_3$ & $u^c_1$ & $u^c_2$ & 
$u^c_3$ & $H_3$ & $h$   & $L_1$ & $L_2$   & $L_3 $  \\
&& $H_4$&$\bar{H}_1$ & $\bar{H}_2$ &&&&&&&&&&&\\             
\hline
\hline
${\cal{D}}_{36}$&   1&-2& 1& 1& 0& 0& 0& 0& 0& 0& 0& 0& 0& 0& 0\\&&
                       0& 0& 0& 0&-1& 2&-1&  &  &  &  &  &  &  \\&&  
                       0& 0& 0& 0& 0& 0& 0& 0& 0& 0& 0& 0& 0& 0\\&& 
                       0& 0& 6& 0& 0& 0&  &  &  &  &  &  &  &  \\&& 
                       0& 0& 0& 0& 0& 0& 0& 0& 0& 0& 0& 0& 0& 0\\&&
                       0& 0& 0&  &  &  &  &  &  &  &  &  &  &  \\
                       
${\cal{D}}_{37}$&   1&-2& 1& 1& 0& 0& 0& 0& 0& 0& 0& 0& 0& 0& 0\\&& 
                       0& 0& 0& 0&-1& 2&-1&  &  &  &  &  &  &  \\&&  
                       0& 0& 0& 0& 0& 0& 0& 0& 0& 0& 0& 0& 0& 0\\&& 
                       0& 0& 0& 6& 0& 0&  &  &  &  &  &  &  &  \\&& 
                       0& 0& 0& 0& 0& 0& 0& 0& 0& 0& 0& 0& 0& 0\\&&
                       0& 0& 0&  &  &  &  &  &  &  &  &  &  &  \\
                       
${\cal{D}}_{38}$&   1& 0&-1& 0& 0& 0& 0& 0& 0& 0& 0& 0& 0& 0& 0\\&&
                       0& 0& 0& 1& 0& 2& 2&  &  &  &  &  &  &  \\&&  
                       0& 0& 0& 0& 0& 0& 0& 0& 0& 0& 0& 0& 0& 0\\&& 
                       0& 0& 0& 0& 0& 0&  &  &  &  &  &  &  &  \\&& 
                       0& 0& 0& 0& 0& 3& 0& 0& 0& 0& 0& 0& 0& 0\\&&
                      -4& 0& 0&  &  &  &  &  &  &  &  &  &  &  \\
                      
${\cal{D}}_{39}$&   1&-2& 1&-2& 0& 0& 0& 0& 0& 0& 0& 0& 0& 0& 0\\&&
                       0& 0& 0& 3&-1& 2& 2&  &  &  &  &  &  &  \\&&  
                       0& 0& 0& 0& 0& 0& 0& 0& 0& 0& 0& 0& 0& 0\\&& 
                       0& 0& 0& 0& 0& 0&  &  &  &  &  &  &  &  \\&& 
                       0& 0& 0& 0& 0& 0& 0& 0& 0& 0& 0& 0& 0& 3\\&&
                      -3& 0& 0&  &  &  &  &  &  &  &  &  &  &  \\
                       
${\cal{D}}_{40}$&   1& 1&-2& 1& 0& 0& 0& 0& 0& 3& 0& 0& 0& 0& 0\\&& 
                       0& 0& 0& 3& 2& 2& 2&  &  &  &  &  &  &  \\&&
                       0& 0& 0& 0& 0& 0& 0& 0& 0& 0& 0& 0& 0& 0\\&& 
                       0& 0& 0& 0& 0& 0&  &  &  &  &  &  &  &  \\&& 
                       0& 0& 0& 0& 0& 0& 0& 0& 0& 0& 0& 0& 0& 0\\&&
                       0& 0& 0&  &  &  &  &  &  &  &  &  &  &  \\
                         
${\cal{D}}_{41}$&   1&-2&-2& 1& 0& 0& 0& 0& 0& 0& 0& 0& 0& 0& 0\\&& 
                       0& 0& 0& 0&-1&-1& 2&  &  &  &  &  &  &  \\&& 
                       0& 0& 0& 0& 0& 0& 0& 0& 6& 0& 0& 0& 0& 0\\&& 
                       0& 0& 0& 0& 0& 0&  &  &  &  &  &  &  &  \\&& 
                       0& 0& 0& 0& 0& 0& 0& 0& 0& 0& 0& 0& 0& 0\\&&
                       0& 0& 0&  &  &  &  &  &  &  &  &  &  &  \\
                       
\hline
\hline
\end{tabular}
Table 3.a continued.
\end{table}
\end{flushleft}

\begin{flushleft}
\begin{table}
\begin{tabular}{|r||r||rrrrrrrrrrrrrr|}
\hline 
\hline
FD &$\frac{Q^{(A)}}{15}$& $\Phi_{46}$ & $\Phi^{'}_{45}$ & $\bar{\Phi}^{'}_{56}$&$V_3$&$V_2$&
$V_6$&$V_5$&$V_9$&$V_8$&$N^{c}_{1}$&$N^{c}_{2}$&$N^{c}_{3}$&$\Phi^{'}_{46}$
      &$\Phi_{45}$\\
& &$\bar{\Phi}_{56}$&$e^c_1$& $e^c_2$& $e^c_3$ &$H_7$& $H_6$ & $H_5$ &&&&&&&\\
& & $\bar{\Phi}^{\alpha\beta}_{1,2}$ & $H_{11}$ & $H_{10}$ & $\bar{H}_{9}$ & 
	$\bar{H}_{8}$ & $H_{16}$ & $H_{14}$ & $H_{12}$ & $V_{1}$ & $V_{4}$ & 
	$V_{7}$ & $H_{17}$ & $H_{15}$ & $H_{13}$ \\
& & $V_{15}$& $V_{14}$ & $V_{13}$ & $V_{12}$ & $V_{10}$ & $V_{11}$&&&&&&&&\\
&& $Q_1$ & $Q_2$ & $Q_3$ & $d^c_1$ & $d^c_2$ & $d^c_3$ & $u^c_1$ & $u^c_2$ &
	$u^c_3$ & $H_3$ & $h$   & $L_1$ & $L_2$   & $L_3 $  \\
&& $H_4$&$\bar{H}_1$ & $\bar{H}_2$ &&&&&&&&&&&\\             
\hline
\hline
${\cal{D}}_{42}$&   1&-2& 1& 4& 0& 0& 0& 0& 0& 0& 0& 0& 0& 0& 0\\&&
                       0& 0& 0& 6&-4&-1& 5&  &  &  &  &  &  &  \\&& 
                       0& 0& 0& 0& 0& 0& 0& 0& 0& 0& 0&12& 0& 0\\&& 
                       0& 0& 0& 0& 0& 0&  &  &  &  &  &  &  &  \\&& 
                       0& 0& 0& 0& 0& 0& 0& 0& 0& 0& 0& 0& 0& 0\\&&
                       0& 0& 0&  &  &  &  &  &  &  &  &  &  &  \\
                        
${\cal{D}}_{43}$&   1&-2& 1&-2& 0& 0& 0& 0& 0& 0& 0& 0& 0& 0& 0\\&& 
                       0& 0& 0& 0& 2&-1&-1&  &  &  &  &  &  &  \\&& 
                       0& 0& 0& 0& 0& 0& 0& 0& 0& 0& 6& 0& 0& 0\\&& 
                       0& 0& 0& 0& 0& 0&  &  &  &  &  &  &  &  \\&& 
                       0& 0& 0& 0& 0& 0& 0& 0& 0& 0& 0& 0& 0& 0\\&&
                       0& 0& 0&  &  &  &  &  &  &  &  &  &  &  \\
                        
${\cal{D}}_{44}$&   1& 1& 1&-2& 0& 0& 0& 0& 0& 0& 0& 0& 0& 0& 0\\&&
                       0& 0& 0& 0&-1&-1& 2&  &  &  &  &  &  &  \\&&
                       0& 0& 0& 0& 0& 0& 0& 0& 0& 0& 0& 0& 0& 0\\&& 
                       0& 0& 0& 0& 0& 6&  &  &  &  &  &  &  &  \\&& 
                       0& 0& 0& 0& 0& 0& 0& 0& 0& 0& 0& 0& 0& 0\\&&
                       0& 0& 0&  &  &  &  &  &  &  &  &  &  &  \\
                        
${\cal{D}}_{45}$&   2&-4&-1&-1& 0& 0& 0& 0& 0& 0& 0& 0& 0& 0& 0\\&& 
                       0& 0& 0& 6& 1& 4& 1&  &  &  &  &  &  &  \\&&  
                       0& 0& 0& 0& 0& 0& 0& 0& 0& 0& 0& 0& 0& 0\\&& 
                       0& 0& 0& 0& 0& 0&  &  &  &  &  &  &  &  \\&& 
                       0& 0& 0& 0& 0& 0& 0& 0& 0& 0& 0& 6& 0& 0\\&&
                      -6& 0& 0&  &  &  &  &  &  &  &  &  &  &  \\
                       
${\cal{D}}_{46}$&   2& 2&-1&-1& 0& 6& 0& 0& 0& 0& 0& 0& 0& 0& 0\\&& 
                       0& 0& 0& 6& 1& 4& 7&  &  &  &  &  &  &  \\&&  
                       0& 0& 0& 0& 0& 0& 0& 0& 0& 0& 0& 0& 0& 0\\&& 
                       0& 0& 0& 0& 0& 0&  &  &  &  &  &  &  &  \\&& 
                       0& 0& 0& 0& 0& 0& 0& 0& 0& 0& 0& 0& 0& 0\\&&
                       0& 0& 0&  &  &  &  &  &  &  &  &  &  &  \\
                       
${\cal{D}}_{47}$&   2&-1&-1& 2& 0& 0& 0& 6& 0& 0& 0& 0& 0& 0& 0\\&& 
                       0& 0& 0& 6& 1& 7& 4&  &  &  &  &  &  &  \\&&
                       0& 0& 0& 0& 0& 0& 0& 0& 0& 0& 0& 0& 0& 0\\&& 
                       0& 0& 0& 0& 0& 0&  &  &  &  &  &  &  &  \\&& 
                       0& 0& 0& 0& 0& 0& 0& 0& 0& 0& 0& 0& 0& 0\\&&
                       0& 0& 0&  &  &  &  &  &  &  &  &  &  &  \\
                         
\hline
\hline
\end{tabular}
Table 3.a continued.
\end{table}
\end{flushleft}

\begin{flushleft}
\begin{table}
\begin{tabular}{|r||r||rrrrrrrrrrrrrr|}
\hline 
\hline
FD &$\frac{Q^{(A)}}{15}$& $\Phi_{46}$ & $\Phi^{'}_{45}$ & $\bar{\Phi}^{'}_{56}$&$V_3$&$V_2$&
$V_6$&$V_5$&$V_9$&$V_8$&$N^{c}_{1}$&$N^{c}_{2}$&$N^{c}_{3}$&$\Phi^{'}_{46}$ &
		$\Phi_{45}$\\
& &$\bar{\Phi}_{56}$&$e^c_1$& $e^c_2$& $e^c_3$ &$H_7$& $H_6$ & $H_5$ &&&&&&&\\
& & $\bar{\Phi}^{\alpha\beta}_{1,2}$ & $H_{11}$ & $H_{10}$ & $\bar{H}_{9}$ & 
	$\bar{H}_{8}$ & $H_{16}$ & $H_{14}$ & $H_{12}$ & $V_{1}$ & $V_{4}$ & 
	$V_{7}$ & $H_{17}$ & $H_{15}$ & $H_{13}$ \\
& & $V_{15}$& $V_{14}$ & $V_{13}$ & $V_{12}$ & $V_{10}$ & $V_{11}$&&&&&&&&\\
&& $Q_1$ & $Q_2$ & $Q_3$ & $d^c_1$ & $d^c_2$ & $d^c_3$ & $u^c_1$ & $u^c_2$ & 
$u^c_3$ & $H_3$ & $h$   & $L_1$ & $L_2$   & $L_3 $  \\
&& $H_4$&$\bar{H}_1$ & $\bar{H}_2$ &&&&&&&&&&&\\                         
\hline
\hline
${\cal{D}}_{48}$&   2& 0& 1&-3& 0& 0& 0& 0& 0& 0& 0& 0& 0& 0& 0\\&& 
                       0& 0& 0& 2& 3& 4& 1&  &  &  &  &  &  &  \\&&  
                       0& 0& 0& 0& 0& 0& 0& 0& 0& 0& 0& 0& 0& 0\\&& 
                       0& 0& 0& 0& 0& 0&  &  &  &  &  &  &  &  \\&& 
                       0& 0& 0& 6& 0& 0& 0& 0& 0& 0& 0& 0& 0& 0\\&&
                      -8& 0& 0&  &  &  &  &  &  &  &  &  &  &  \\
                       
${\cal{D}}_{49}$&   2&-3& 1& 0& 0& 0& 0& 0& 0& 0& 0& 0& 0& 0& 0\\&& 
                       0& 0& 0& 2& 3& 1& 4&  &  &  &  &  &  &  \\&&
                       0& 0& 0& 0& 0& 0& 0& 0& 0& 0& 0& 0& 0& 0\\&& 
                       0& 0& 0& 0& 0& 0&  &  &  &  &  &  &  &  \\&&
                       0& 0& 0& 0& 6& 0& 0& 0& 0& 0& 0& 0& 0& 0\\&&
                      -8& 0& 0&  &  &  &  &  &  &  &  &  &  &  \\
                         
${\cal{D}}_{50}$&   2&-1&-1&-4& 0& 0& 0& 0& 0& 0& 0& 0& 0& 0& 0\\&& 
                       0& 0& 0& 6& 1& 1& 4&  &  &  &  &  &  &  \\&&
                       0& 0& 0& 0& 0& 0& 0& 0& 0& 0& 0& 0& 0& 0\\&& 
                       0& 0& 0& 0& 0& 0&  &  &  &  &  &  &  &  \\&& 
                       0& 0& 0& 0& 0& 0& 0& 0& 0& 0& 0& 0& 6& 0\\&&
                      -6& 0& 0&  &  &  &  &  &  &  &  &  &  &  \\
                         
\hline
\hline
\end{tabular}
Table 3.a continued.
\end{table}
\end{flushleft}
\hfill\vfill\newpage

\begin{flushleft}
\begin{table}
\begin{tabular}{|r|r|||r|r|||r|r|||r|r|||r|r|}
\hline 
\hline
FD & VEV & FD & VEV & FD & VEV & FD & VEV & FD & VEV\\
\hline
\hline
\hline
\bda{1} & $N^{c}_{1}$  & \bda{11}& $\bar{H}_{2}$  & \bda{21}& ${u}^{c}_{3}$  &
	 \bda{31}& $V_{ 4}$  & \bda{41}&$V_{1}$\\
\bda{2} & $N^{c}_{2}$  & \bda{12}& $\bar{H}_{1}$  & \bda{22}& $h$            &
	 \bda{32}& $H_{13}$  & \bda{42}&$H_{17}$\\
\bda{3} & $N^{c}_{3}$  & \bda{13}& $H_{10}$       & \bda{23}& ${e}^{c}_{2}$  &
	 \bda{33}& $V_{10}$  & \bda{43}&$V_{7}$\\
\bda{4} & $H_{16}$     & \bda{14}& $\Phi^{'}_{46}$  
                                                  & \bda{24}& ${e}^{c}_{1}$  &
	 \bda{34}& $V_{15}$  & \bda{44}&$V_{11}$\\
\bda{5} & $H_{14}$     & \bda{15}& $\bar{\Phi}_{56}$
                                                  & \bda{25}& $V_{ 9}$       &
	 \bda{35}& $V_{14}$  & \bda{45}&$L_{1}$\\
\bda{6} & $H_{12}$     & \bda{16}& $\bar{\Phi}^{\alpha\beta}_{1,2}$ 
                                                  & \bda{26}& $V_{ 3}$       &
	 \bda{36}& $V_{13}$  & \bda{46}&$V_{2}$\\
\bda{7} & $H_{ 3}$     & \bda{17}& $\bar{H}_{8}$  & \bda{27}& ${u}^{c}_{1}$  &
	 \bda{37}& $V_{12}$  & \bda{47}&$V_{5}$\\
\bda{8} & $\bar{H}_{9}$& \bda{18}& $V_{ 6}$       & \bda{28}& $Q_{ 2}$       &
	 \bda{38}& ${d}^{c}_{3}$ &\bda{48}&${d}^{c}_{1}$\\
\bda{9} & $H_{11}$     & \bda{19}& $Q_{ 1}$       & \bda{29}& ${u}^{c}_{2}$  &
	 \bda{39}& ${L}_{3}$  &\bda{49}&${d}^{c}_{2}$\\
\bda{10}& $\Phi_{45}$  & \bda{20}& $Q_{3}$  & \bda{30}& $H_{15}$       &
		 \bda{40}& $V_{ 8}$  & \bda{50}&$L_{2}$\\

\hline
\end{tabular}
\\
Table 3.b. Unique VEV associated with each $D$-Flat basis direction.
\end{table}
\end{flushleft}

\end{document}